\documentclass[12pt,draftclsnofoot,onecolumn]{IEEEtran}
\usepackage{amsfonts}
\usepackage{amsmath}
\usepackage{amssymb}
\usepackage{bm}
\usepackage{xcolor,graphicx,float}
\usepackage{color, soul}
\usepackage{stfloats}
\usepackage[numbers,sort&compress]{natbib}
\usepackage[amsmath,thmmarks]{ntheorem}
\usepackage{theorem}
\usepackage{algorithm}
\usepackage{algorithmic}
\usepackage{graphicx}
\usepackage{subfigure}
\usepackage{pict2e}

\newtheorem{lemma}{Lemma}

\theoremheaderfont{\sc}\theorembodyfont{\upshape}
\theoremstyle{nonumberplain}
\theoremseparator{}
\theoremsymbol{\rule{1ex}{1ex}}

\begin{document}
\title{Gridless Multisnapshot Variational Line Spectral Estimation from Coarsely Quantized Samples}

\author{Ning Zhang, Jiang~Zhu and Zhiwei Xu \thanks{Ning Zhang is with the Nanjing Marine Radar Institute, Nanjing, China. Jiang Zhu and Zhiwei Xu are with the Ocean College,
Zhejiang University, No.1 Zheda Road, Zhoushan, 316021, China.}}
\maketitle
\begin{abstract}
Due to the increasing demand for low power and higher sampling rates, low resolution quantization for data acquisition has drawn great attention recently. Consequently, line spectral estimation (LSE) with multiple measurement vectors (MMVs) from coarsely quantized samples is of vital importance in cutting edge array signal processing applications such as range estimation and DOA estimation in millimeter wave radar systems. In this paper, we combine the low complexity gridless variational line spectral estimation (VALSE) and expectation propagation (EP) and propose an MVALSE-EP algorithm to estimate the frequencies from coarsely quantized samples. In addition, the Cram\'{e}r Rao bound (CRB) is derived as a benchmark performance of the proposed algorithm, and insights are provided to reveal the effects of system parameters on estimation performance. It is shown that snapshots benefits the frequency estimation, especially in coarsely quantized scenarios. Numerical experiments are conducted to demonstrate the effectiveness of MVALSE-EP, including real data set.
\end{abstract}
{\bf keywords}: line spectral estimation, multiple measurement vectors, quantization, variational Bayes, expectation propagation
\section{Introduction}
As radar and communication systems scale up in bandwidth, the cost and power consumption of high-precision (e.g., $10-12$ bits) analog-to-digital converters (ADCs) become the limiting factor for various applications, such as cognitive radio and radar. As a remedy, low resolution quantization has been gradually drawn attention in recent years \cite{CTD09Phd, FangAD, Lijian19Taes, Lijian17Asic, Lijian19TSP, Zhu20TAES, GL19ICASSP,AW21ICASSP}. In addition, in military radar systems incorporating very large arrays, low resolution quantization is a promising technique as it provides a reasonable estimation accuracy, and it also enables a very high sampling rate for a wideband system. Moreover, it dramatically reduces the memory and transmission requirements for the samples.

For low resolution quantization, the loss of information per sample is large, and new, low-resolution quantization adapted digital signal processing algorithms should be developed \cite{radarclutter, RFImitigation}. From the linear signal processing point of view, the spectrum of 1-bit line spectral signal is analyzed. As shown in \cite{SAR1991,Zhu20TAES}, the binary data contains plentiful self-generated \cite{SAR1991} and cross-generated harmonics \cite{Zhu20TAES}. For low signal to noise (SNR) scenario, the strengths of harmonics decay quickly and conventional fast Fourier transform (FFT) performs well. While for high SNR scenario, FFT will overestimate the model order. From the nonlinear signal processing point of view, the direction of arrival (DOA) estimation from 1 bit data is treated as a nonlinear parameter estimation problem. In \cite{SingletoneTSP}, the frequency estimation from 1 bit quantized samples has been addressed. The problems in \cite{SingletoneTSP} adopts the deterministic signal modeling by treating the signal samples as a deterministic, yet unknown set of parameters that has to be estimated. By comparing the derived Cram\`{e}r-Rao bound (CRB) for 1-bit quantization with that of infinite quantization, it is found that 1 bit quantization gives a dramatic increase of variance at certain frequencies, and a slightly worse performance for other frequencies. In \cite{DOA1bit02}, a single stochastic Gaussian point source model is assumed to study the DOA estimation which is a spatial analogy to the problem of temporal line spectral estimation (LSE) with multiple measurement vectors (MMVs), and the CRB for a two-sensor array case is derived. It is shown that the estimation error has weak dependency on signal-to-noise ratio (SNR), and there exist two singular DOA angles $0^{\circ}$ and $30^{\circ}$ for which higher SNR results in better estimation performance.

In this work, we present an approach named multisnapshot VALSE-EP (MVALSE-EP) for LSE with MMVs from coarsely quantized measurements. Compared to  \cite{DOA1bit02}, we use a deterministic signal model by treating the amplitudes of the signal as unknown deterministic parameters, instead of stochastic Gaussian model. For the stochastic signal model, the spatial signal samples are clearly dependent (as the sensors are correlated), and the covariance matrix is not diagonal. In order to apply the \emph{arcsine} law for LSE, Gaussian assumption for the signal amplitude must be adopted. This implies that when the signal amplitude distribution violates the Gaussian assumption, the estimation performance of the covariance matrix based approach may degrade significantly. Besides, the number of snapshots is usually needed to be large to obtain accurate reconstruction of the covariance matrix, e.g., $10^4$.  By using the deterministic signal model, the signal samples can be regarded as independent identically distributed (i.i.d.) variables, and consequently, performing approximate Bayesian inference becomes tractable \cite{Minka, UnifiedSPL, VALSEEP}, as we show later. Compared to \cite{VALSEEP} which proposes the VALSE-EP to tackle the LSE in the single snapshot case, we extend it to address the multisnapshot LSEs. This is meaningful especially in coarsely quantized scenarios as model may become unidentifiable in the single snapshot case in certain scenarios \cite{JiayingTSP19}, while multisnapshots makes the estimation problem unlikely identifiable and benefits the estimation performance \cite{Yangmultichannel}.

In this paper, the LSE estimation with MMVs from coarsely quantized samples is studied. Firstly, to provide a bench mark performance of the LSE, the CRB is derived. It is shown that under $1$ bit quantization, the CRB is inversely proportional to the number of snapshots and the cubic of the number of antennas. For lower SNR scenario, the CRB is inversely proportional to SNR, while the CRB is inversely proportional to the square root of the SNR for high SNR scenario. Secondly, the MVALSE-EP is proposed to estimate the DOAs and the complex weight. Since MVALSE-EP performs Newton step to refine the frequencies, it overcomes the model mismatch \cite{Modelmismatch} issue incurred by on-grid assumptions \cite{Yangzaireview}. The expectation maximization (EM) algorithm is incorporated to estimate the noise variance automatically. Finally, numerical experiments are conducted to demonstrate the effectiveness of MVALSE-EP, including real data set.

The rest of this paper is organized as below. Section \ref{setup} describes the system model. Section \ref{bound} derives CRB. The MVALSE-EP algorithm and the details of the updating expressions are presented in Section \ref{Algorithm}. Substantial numerical experiments are provided in Section \ref{NS} and Section \ref{con} concludes the paper.

For a complex vector ${\mathbf x}\in {\mathbb C}^M$, let ${\Re}\{{\mathbf x}\}$ and $\Im\{\mathbf x\}$ denote the real and imaginary part of $\mathbf x$, respectively, let $|\mathbf x|$ and $\angle{\mathbf x}$ denote the componentwise amplitude and phase of $\mathbf x$, respectively. For the square matrix $\mathbf A$, let ${\rm diag}(\mathbf A)$ return a vector with elements being the diagonal of $\mathbf A$. While for a vector ${\mathbf a}$, let ${\rm diag}(\mathbf a)$ return a diagonal matrix with the diagonal being $\mathbf a$, and thus ${\rm diag}({\rm diag}(\mathbf A))$ returns a diagonal matrix. Let ${\rm j}$ denote the imaginary number. Let ${\mathcal S}\subset \{1,\cdots,N\}$ be a subset of indices and $|{\mathcal S}|$ denote its cardinality. For the matrix $\mathbf J\in\mathbb C^{N\times N}$, let $\mathbf J_{\mathcal S}$ denote the submatrix by choosing both the rows and columns of $\mathbf J$ indexed by $\mathcal S$. Similarly, let $\mathbf h_{\mathcal S}$ denote the subvector by choosing the elements of $\mathbf h$ indexed by $\mathcal S$. Let ${(\cdot)}^{*}_{\mathcal S}$, ${(\cdot)}^{\rm T}_{\mathcal S}$ and ${(\cdot)}^{\rm H}_{\mathcal S}$ be the conjugate, transpose and Hermitian transpose operator of ${(\cdot)}_{\mathcal S}$, respectively. For the matrix $\mathbf A$, let $|\mathbf A|$ denote the elementwise absolute value of $\mathbf A$. Let $\mathbf I_L$ denote the identity matrix of dimension $L$. $``\sim i"$ denotes the indices ${\mathcal S}$ excluding $i$. Let ${\mathcal {CN}}({\mathbf x};{\boldsymbol \mu},{\boldsymbol \Sigma})$ denote the complex normal (CN) distribution of ${\mathbf x}$ with mean ${\boldsymbol \mu}$ and covariance ${\boldsymbol \Sigma}$. For a frequency $\theta$, let ${\mathcal VM}(\theta;\mu,\kappa)$ be the von Mises distribution with $\mu$ and $\kappa$ being the mean and concentration parameters. For a random vector $\mathbf x$ with probability density function (PDF) $p({\mathbf x})$, let ${\rm Proj}[p({\mathbf x})]$ denote the projection of $p({\mathbf x})$ onto Gaussian PDF with diagonal covariance matrix, where the means and variances are matched with that of $p({\mathbf x})$. Let $\phi(x)=\exp(-x^2/2)/{\sqrt{2\pi}}$ and $\Phi(x)=\int_{-\infty}^x\phi(t){\rm d}t$ denote the standard normal probability density function (PDF) and cumulative distribution function (CDF), respectively.

\section{Problem Setup}\label{setup}
Consider a line spectral estimation problem with MMVs. For the $t$th snapshot, the signal received before quantization can be modeled as
\begin{align}\label{modelorig}
{\mathbf r}(t)={\mathbf A}({\boldsymbol \omega}){\mathbf x}(t)+{\mathbf w}(t),\quad t=1,2,\cdots,T,
\end{align}
where ${\mathbf x}(t)=[x_{1}(t),x_{2}(t),\cdots,x_{K}(t)]^{\rm T}$ denotes the complex coefficients of the unknown frequencies ${\boldsymbol \omega}=[\omega_1,\omega_2,\cdots,\omega_K]^{\rm T}\in {\mathbb R}^K$, $\omega_i\in[0,2\pi)$, and $K$ denotes the number of frequencies,
\begin{align}
{\mathbf A}=[{\mathbf a}(\omega_1),{\mathbf a}(\omega_2),\cdots,{\mathbf a}(\omega_k),\cdots,{\mathbf a}(\omega_K)]\in {\mathbb C}^{N\times K}
\end{align}
denotes the nominal array manifold matrix, with the steering vectors ${\mathbf a}(\omega_k)=[1,{\rm e}^{{\rm j}\omega_k},\cdots,{\rm e}^{{\rm j}(N-1)\omega_k}]^{\rm T}$ as its columns, ${\mathbf w}(t) \in {\mathbb C}^M$ denotes the additive internal (e.g., thermal) receiver noise modeled as spatially and temporally independent, identically distributed (i.i.d.) zero-mean circular CN with a covariance matrix ${\mathbf R}_w\triangleq {\rm E}[{\mathbf w}(t){\mathbf w}^{\rm H}(t)]=\sigma^2{\mathbf I}_N$ where $\sigma^2$ is unknown.


The signal ${\mathbf r}(t)$ is then coarsely quantized to obtain
\begin{align}\label{quantmodel}
{\mathbf y}(t)={\mathcal Q}\left\{\Re\left\{{\mathbf r}(t) \right\}\right\}+{\rm j}{\mathcal Q}\left\{\Im\left\{{\mathbf r}(t) \right\}\right\},\quad t=1,2,\cdots,T,
\end{align}
where ${\mathcal Q}(\cdot)$ is a quantizer which maps the continuous-valued observations into a finite number of bits. Note that for a quantizer with bit-depth $B$, the cardinality of the output of the quantizer is $|{\mathcal D}|=2^B$. Assume that the quantization intervals for the quantizer ${\mathcal Q}(\cdot)$ are $\{(\tau_l,\tau_{l+1})\}_{l=0}^{|{\mathcal D}|-1}$, where $\tau_0=-\infty$, $\tau_{|{\mathcal D}|}=\infty$, $\bigcup_{l=0}^{|{\mathcal D}|-1}[\tau_l,\tau_{l+1})={\mathbb R}$. Given a real number $a\in [\tau_l,\tau_{l+1})$ input, the quantizer output is
\begin{align}
{\mathcal Q}(a)=o_l.
\end{align}
For example, one-bit quantization refers to $B=1$, $D=2$, $\tau_0=-\infty$, $\tau_1=0$ and $\tau_{2}=\infty$, ${\mathcal Q}(\cdot)$ reduces to the signum function, i.e., $Q(\cdot)={\rm sign}(\cdot)$.

The SNR of the $k$th target is defined as
\begin{align}
{\rm SNR}_k=10\log_{10}\frac{\sum\limits_{t=1}^T|x_k(t)|^2/T}{\sigma^2},
\end{align}
which corresponds to the sample SNR. Note that after coherent integration, the sample SNR can be improved up to $10\log_{10}N$ dB.

It is worth noting that model (\ref{modelorig}) and model (\ref{quantmodel}) has various applications such as range, velocity and angle estimation in millimeter wave LFMCW radar \cite{LFMCWfast}. Below we present the details.
\begin{itemize}
  \item range estimation: here $\omega_{k}=2\pi\times\frac{2 \mu T_s}{c}r_k$ where $\mu$, $T_s$, $c$ and $r_k$ denote the chirp rate, sampling interval, electromagnetic speed and radial distance of the $k$th target.
  \item velocity estimation: here $\omega_{k}=2\pi\times \frac{2T_c}{\lambda}v_k$ where $T_c$, $\lambda$ and $v_k$ denote the repetition interval, wavelength and radial velocity of the $k$th target.
  \item DOA estimation: here $\omega_{k}=2\pi \times \frac{d}{\lambda}\sin\theta_k$ where $d$, $\lambda$ and $\theta_k$, denote the element spacing, wavelength and direction of arrival of the $k$th target.
  \item vital signs detection \cite{VDS}:  The vital sign extraction from multiple independent self-injection locking doppler radars can be formulated as a line spectral estimation problem with MMVs, see \cite[eq. (5)]{VDS}.
\end{itemize}

\section{Cram\'{e}r Rao bound}\label{bound}
The CRB is a lower bound of unbiased estimators and provides a benchmark against which we can compare the performance of the proposed algorithm \cite{Kay, peterstocia}. To derive the CRB, $K$ is assumed to be known, the frequencies ${\boldsymbol \omega}\in{\mathbb R}^K$ and weights ${\mathbf X}=[{\mathbf x}(1),\cdots,{\mathbf x}(T)]\in{C}^{K\times T}$ are treated as unknown deterministic parameters. As for the quantizer $Q(\cdot)$, the quantization intervals are $\{(\tau_l,\tau_{l+1})\}_{l=0}^{|{\mathcal D}|-1}$, where $\tau_0=-\infty$, $\tau_{{\mathcal D}}=\infty$, $\bigcup_{l=0}^{{\mathcal D}-1}[\tau_l,\tau_{l+1})={\mathbb R}$. Given a real number $a\in [\tau_l,\tau_{l+1})$, the representation is
\begin{align}
Q(a)=o_l, \quad {\rm if}\quad a\in [\tau_l,\tau_{l+1}).
\end{align}
Note that for a quantizer with bit-depth $B$, the cardinality of the output of the quantizer is $|{\mathcal D}|=2^B$.

The CRB equals to the inverse of the Fisher information matrix (FIM). For calculating the FIM, the following lemma can be utilized.
\begin{lemma}\label{lemmaFIM}
\cite{Fu, VALSEEP} Let ${\boldsymbol \kappa}\in {\mathbb R}^P$ denote the set of unknown deterministic parameters. Note that in the case of quantized observations ${\mathbf y}={\mathcal Q}({\mathbf r})\in{\mathbb R}^N$ where ${\mathbf r}\sim {\mathcal {N}}({\boldsymbol \mu}({\boldsymbol \kappa}),{\sigma}^2{\mathbf I}_N/2)$, the FIM is given by
\begin{align}\label{FIMcal}
{\mathbf I}({\boldsymbol \kappa})=\frac{2}{\sigma^2}\left[\frac{\partial {\boldsymbol \mu}({\boldsymbol \kappa})}{\partial {\boldsymbol \kappa}^{\rm T}}\right]^{\rm T}{\boldsymbol \Lambda}\left[\frac{\partial {\boldsymbol \mu}({\boldsymbol \kappa})}{\partial {\boldsymbol \kappa}^{\rm T}}\right],
\end{align}
where
\begin{align}
\frac{\partial {\boldsymbol \mu}({\boldsymbol \kappa})}{\partial {\boldsymbol \kappa}^{\rm T}}=\left[
                                                                                         \begin{array}{cccc}
                                                                                           \frac{\partial [{\boldsymbol \mu}({\boldsymbol \kappa})]}{\partial \kappa_1} & \frac{\partial [{\boldsymbol \mu}({\boldsymbol \kappa})]}{\partial \kappa_2} & \cdots & \frac{\partial [{\boldsymbol \mu}({\boldsymbol \kappa})]}{\partial \kappa_P}
                                                                                         \end{array}
                                                                                       \right]\in {\mathbb R}^{N\times P},
\end{align}
and ${\boldsymbol \Lambda}$ is a diagonal matrix with the $(i,i)$th element
\begin{align}
\Lambda_{i,i}=h(\mu_i(\boldsymbol \kappa),\sigma^2),
\end{align}
and $h(x,\sigma^2)$ is
\begin{align}
h(x,\sigma^2)=\sum\limits_{d=0}^{|{\mathcal D}|-1}\frac{\left[\phi\left(\frac{\tau_{d+1}-x}{\sigma/\sqrt{2}}\right)-\phi\left(\frac{\tau_{d}-x}{\sigma/\sqrt{2}}\right)\right]^2}{\Phi\left(\frac{\tau_{d+1}-x}{\sigma/\sqrt{2}}\right)-\Phi\left(\frac{\tau_{d}-x}{\sigma/\sqrt{2}}\right)}.
\end{align}
For one-bit quantization where $|{\mathcal D}|=2$, $\tau_0=-\infty$, $\tau_1=0$, $\tau_2=\infty$, $h(x,\sigma^2)$ simplifies to be
\begin{align}\label{onebith}
h(x,\sigma^2)=\frac{\phi^2\left(\frac{x}{\sigma/\sqrt{2}}\right)}{\Phi\left(\frac{x}{\sigma/\sqrt{2}}\right)\Phi\left(\frac{-x}{\sigma/\sqrt{2}}\right)}=\frac{1}{2\pi}\frac{{\rm e}^{-\frac{2x^2}{\sigma^2}}}{\Phi\left(\frac{x}{\sigma/\sqrt{2}}\right)\Phi\left(\frac{-x}{\sigma/\sqrt{2}}\right)}.
\end{align}
\end{lemma}
For unquantized system, the FIM (\ref{FIMcal}) is obtained with ${\boldsymbol \Lambda}={\mathbf I}_N$.

In our setting, the observations are \small{$\left[\Re\{{\mathbf y}(1)\};\Im\{{\mathbf y}(1)\};\Re\{{\mathbf y}(2)\};\Im\{{\mathbf y}(2)\};\cdots;\Re\{{\mathbf y}(T)\};  \Im\{{\mathbf y}(T)\} \right]$}. Note that ${\boldsymbol \kappa}\in{\mathbb R}^{2KT+K}$ and ${\boldsymbol \mu}({\boldsymbol \kappa})\in{\mathbb R}^{2MT}$ are
\begin{align}\label{partres}
{\boldsymbol \kappa}=\left[\begin{array}{c}
  \Re\{{\mathbf x}(1)\} \\
  \Im\{{\mathbf x}(1)\} \\
  \Re\{{\mathbf x}(2)\} \\
  \Im\{{\mathbf x}(2)\} \\
  \vdots\\
  \vdots\\
  \Re\{{\mathbf x}(T)\} \\
  \Im\{{\mathbf x}(T)\} \\
  {\boldsymbol \omega}
\end{array}
 \right],
 {\boldsymbol \mu}({\boldsymbol \kappa})=\left[\begin{array}{c}
  \Re\{{\mathbf A}{\mathbf x}(1)\} \\
  \Im\{{\mathbf A}{\mathbf x}(1)\} \\
  \Re\{{\mathbf A}{\mathbf x}(2)\} \\
  \Im\{{\mathbf A}{\mathbf x}(2)\} \\
  \vdots\\
  \vdots\\
  \Re\{{\mathbf A}{\mathbf x}(T)\} \\
  \Im\{{\mathbf A}{\mathbf x}(T)\}
\end{array}
 \right].
\end{align}
Define
\begin{subequations}
\begin{align}
\bar{\mathbf A}&=\left[
                  \begin{array}{cc}
                    \Re\{{\mathbf A}\} & -\Im\{{\mathbf A}\} \\
                    \Im\{{\mathbf A}\} & \Re\{{\mathbf A}\} \\
                  \end{array}
                \right],\label{Abar}\\
                &{\mathbf D}_{\omega}=\left[\frac{{\rm d} {\mathbf a}({\omega}_{1})}{{\rm d}{\omega}_{1}},\frac{{\rm d} {\mathbf a}({\omega}_{2})}{{\rm d}{\omega}_{2}},\cdots,\frac{{\rm d} {\mathbf a}({\omega}_{K})}{{\rm d}{\omega}_{K}}\right],\label{Domega}\\
                &{\mathbf X}(t)\triangleq {\rm diag}({\mathbf x}(t)),\label{Xt}\\
&\bar{\mathbf D}(t)=\left[
                  \begin{array}{cc}
                  \frac{\partial{\Re\{{\mathbf {Ax}}(t)\}}}{\partial {\boldsymbol \omega}}
                     \\
                    \frac{\partial{\Im\{{\mathbf {Ax}}(t)\}}}{\partial {\boldsymbol \omega}}
                  \end{array}
                \right]=\left[
                  \begin{array}{cc}
                  \Re\{{\mathbf D}_{\omega}{\mathbf X}(t)\}
                  \\
                  \Im\{{\mathbf D}_{\omega}{\mathbf X}(t)\}
                  \end{array}
                \right],\label{Dbart}\\
&{\boldsymbol\lambda}(t)=\left[\begin{array}{c}
             h(\Re\{{\mathbf A}{\mathbf x}(t)\},{\sigma}^2) \\
             h(\Im\{{\mathbf A}{\mathbf x}(t)\},{\sigma}^2)
           \end{array}\right],\\
          &{\boldsymbol\Lambda}(t)={\rm diag}({\boldsymbol\lambda(t)}),\\
          &{\boldsymbol\Lambda}=
\begin{pmatrix}
  {\boldsymbol\Lambda}(1) &  & &   \\
   & {\boldsymbol\Lambda}(2)& &    \\
   &  & \ddots &     \\
   & &  & {\boldsymbol\Lambda}(T)  \end{pmatrix},\\
   &\bar{\mathbf H}(t)=\bar{\mathbf A}^{\rm T}{\boldsymbol\Lambda}(t)\bar{\mathbf A},\label{Hbart}\\
&\bar{\boldsymbol \Delta}(t)=\bar{\mathbf A}^{\rm T}{\boldsymbol\Lambda}(t)\bar{\mathbf D}(t)\label{Deltabart}.
\end{align}
\end{subequations}
As shown in Appendix \ref{generalFIM}, according to Lemma \ref{lemmaFIM}, the ${\rm CRB}({\boldsymbol \omega})$ is
\begin{align}\label{CRBinv}
{\rm CRB}({\boldsymbol \omega})=\left(\frac{2}{\sigma^2}\sum\limits_{t=1}^T\left(\bar{\mathbf D}^{\rm T}(t){\boldsymbol\Lambda}(t)\bar{\mathbf D}(t)-\bar{\boldsymbol \Delta}^{\rm T}(t)\bar{\mathbf H}^{-1}(t)\bar{\boldsymbol \Delta}({t})\right)\right)^{-1}.
\end{align}

Note that for the DOA estimation problem where $\omega=\frac{2\pi d}{\lambda}\sin\theta$ with $d$ and $\lambda$ being the element spacing and wavelength, by using vector parameter CRB for transformations \cite[eq. (3.30)]{Kay}, the CRB of DOAs is
\begin{align}
{\rm CRB}({\boldsymbol \theta})={\rm diag}\left(\left(\frac{\lambda}{2\pi d\cos({\boldsymbol \theta})}\right)^2\right){\rm CRB}({\boldsymbol \omega}).
\end{align}

We now hope to provide insight into the relationship between the system parameters and the estimation performance. As a result, we provide an asymptotic analysis of the CRB of a single target under one bit quantization for both low SNR and high SNR scenarios. It is assumed that the amplitude of the frequency $g=|x(t)|$ is known and is the same for all the snapshots. Besides, the phase $\psi(t)=\angle x(t)$ of the frequency at the $t$th snapshot is unknown. We assume that $\psi(t)$ is uniformly drawn from $[0,2\pi)$. Define ${\mathbf n}=[0,1,\cdots,N-1]^{\rm T}$. Similar to the derivation of the CRB, here the unknown deterministic parameters ${\boldsymbol \kappa}$ and ${\boldsymbol \mu}({\boldsymbol \kappa})$ are
\begin{align}\label{partressingle}
{\boldsymbol \kappa}=\left[\begin{array}{c}
  \psi(1) \\
  \psi(2) \\
  \vdots\\
  \psi(T-1) \\
  \psi(T) \\
  \omega
\end{array}
 \right],
 {\boldsymbol \mu}({\boldsymbol \kappa})=\left[\begin{array}{c}
  g\cos(\omega{\mathbf n}+\psi(1)) \\
  g\sin(\omega{\mathbf n}+\psi(1))\\
  g\cos(\omega{\mathbf n}+\psi(2)) \\
  g\sin(\omega{\mathbf n}+\psi(2))\\
  \vdots\\
  \vdots\\
  g\cos(\omega{\mathbf n}+\psi(T)) \\
  g\sin(\omega{\mathbf n}+\psi(T))\\
\end{array}
 \right].
\end{align}
 Then
 \begin{align}\label{derpartres}
\frac{\partial {\boldsymbol \mu}({\boldsymbol \kappa})}{\partial {\boldsymbol \kappa}^{\rm T}}=
\begin{pmatrix}
  \bar{\mathbf a}(1) &  & &  &  \bar{\mathbf d}(1) \\
   & \bar{\mathbf a}(2) & &   & \bar{\mathbf d}(2) \\
   &  & \ddots &   & \vdots  \\
   & &  & \bar{\mathbf a}(T)& \bar{\mathbf d}(T)  \end{pmatrix}\in {\mathbb R}^{2NT\times (T+1)}
\end{align}
where $\bar{\mathbf a}(t)$ and $\bar{\mathbf d}(t)$ are defined as
\begin{align}
&\bar{\mathbf a}(t)=\left[
                  \begin{array}{cc}
                  -g\sin({\mathbf n}\omega+\psi(t))
                     \\
                    g\cos({\mathbf n}\omega+\psi(t))
                  \end{array}
                \right],\\
&\bar{\mathbf d}(t)=\left[
                  \begin{array}{cc}
                  \frac{\partial{\Re\{{\mathbf {ax}}(t)\}}}{\partial {\boldsymbol \omega}}
                     \\
                    \frac{\partial{\Im\{{\mathbf {ax}}(t)\}}}{\partial {\boldsymbol \omega}}
                  \end{array}
                \right]=g\left[
                  \begin{array}{cc}
                  -{\mathbf n}\odot \sin({\mathbf n}\omega+\psi(t))
                  \\
                  {\mathbf n}\odot \cos({\mathbf n}\omega+\psi(t))
                  \end{array}
                \right],
\end{align}
respectively, $\odot$ denotes the Hardarmard product, ${\boldsymbol \lambda}(t)$ is
\begin{align}
               {\boldsymbol \lambda}(t)=\left[
                  \begin{array}{cc}
                  h(g\cos({\mathbf n}\omega+\psi(t)),\sigma^2)
                     \\
                    h(g\sin({\mathbf n}\omega+\psi(t)),\sigma^2)
                  \end{array}
                \right]
\end{align}
with $h(\cdot,\cdot)$ being given by (\ref{onebith}). It can be shown that the CRB for a single frequency is
\begin{align}\label{CRBinvsingle}
{\rm CRB}^{-1}({\omega})=\frac{2}{\sigma^2}\sum\limits_{t=1}^T\left(\bar{\mathbf d}^{\rm T}(t){\boldsymbol\Lambda}(t)\bar{\mathbf d}(t)-\frac{\bar{\Delta}^2(t)}{\bar{H}(t)}\right),
\end{align}
where $\bar{\Delta}(t)$ and $\bar{H}(t)$ are
\begin{align}
&\bar{\Delta}(t)=\bar{\mathbf a}^{\rm T}(t){\rm diag}({\boldsymbol \lambda}(t))\bar{\mathbf d}(t)=g^2\sum\limits_{n=0}^{N-1}\chi\left(n,\omega,\psi,\frac{g}{\sigma}\right)n,\\
&\bar{H}(t)=\bar{\mathbf a}^{\rm T}(t){\rm diag}({\boldsymbol \lambda}(t))\bar{\mathbf a}(t)=g^2\sum\limits_{n=0}^{N-1}\chi\left(n,\omega,\psi,\frac{g}{\sigma}\right),
\end{align}
where
\begin{align}
\chi(n,\omega,\psi,\frac{g}{\sigma})=\sin^2(n\omega+\phi(t))h(g\cos(n\omega+\psi(t)),\sigma^2)+\cos^2(n\omega+\psi(t))h(g\sin(n\omega+\psi(t)),\sigma^2).
\end{align}

Now suppose that $\psi(t)\sim {\mathcal U}(0,2\pi)$. We define the expected ${\rm CRB}^{-1}({\omega})$ as
\begin{align}\label{ECRBinv}
{\rm ECRB}^{-1}({\omega})=\frac{2}{\sigma^2}T{\rm E}_{\psi}\left[\bar{\mathbf d}^{\rm T}{\boldsymbol\Lambda}\bar{\mathbf d}-\frac{\bar{\Delta}^2}{\bar{H}}\right].
\end{align}
It can be seen that given that the number of snapshots is large, ${\rm ECRB}^{-1}({\omega})$ (\ref{ECRBinv}) provides a good approximation for ${\rm CRB}^{-1}({\omega})$ (\ref{CRBinvsingle}). Note that for unquantized measurements, one has $\chi(n,\omega,\phi,\frac{g}{\sigma})=1$. It is straightforward to show that ${\rm ECRB}_{\infty}({\omega})$ for a single frequency under unquantized measurements is
\begin{align}
{\rm ECRB}_{\infty}({\omega})= \frac{\sigma^2}{|g|^2}\frac{6}{N(N-1)(N+1)}\frac{1}{T}.
\end{align}
Another property is that the integration
\begin{align}
{\rm E}_{\phi}\left[\chi(n,\omega,\psi,\frac{g}{\sigma})\right]={\rm E}_{\psi}\left[\sin^2(\psi)h(g\cos(\psi),\sigma^2)+\cos^2(\psi)h(g\sin(\psi),\sigma^2)\right]\triangleq r\left(\frac{g}{\sigma}\right)
\end{align}
holds, i,e, the integration is independent of $n\omega$.

Note that
\begin{align}\label{dTd}
\frac{2}{\sigma^2}{\rm E}_{\phi}\left[\bar{\mathbf d}^{\rm T}{\boldsymbol\Lambda}\bar{\mathbf d}\right]&=\frac{2|g|^2}{\sigma^2}\sum\limits_{n=0}^{N-1}n^2{\rm E}_{\phi}\left[\chi(n,\omega,\phi,\frac{g}{\sigma})\right]=\frac{2|g|^2}{\sigma^2}\left(\sum\limits_{n=0}^{N-1}n^2\right)r\left(\frac{g}{\sigma}\right)
\end{align}

It is hard to directly calculate $\frac{2}{\sigma^2}{\rm E}_{\psi}\left[\frac{\bar{\Delta}^2}{\bar{H}}\right]$. Instead we use the Jensen inequality to obtain a bound
\begin{align}\label{jessbound}
\frac{2}{\sigma^2}{\rm E}_{\psi}\left[\frac{\bar{\Delta}^2}{\bar{H}}\right]\geq \frac{2}{\sigma^2}\frac{{\rm E}_{\psi}^2[\bar{\Delta}]}{{\rm E}_{\psi}[\bar{H}]}.
\end{align}
Here
\begin{align}
{\rm E}_{\psi}[\bar{H}]=g^2\sum\limits_{n=0}^{N-1}{\rm E}\left[\chi(n,\omega,\psi,\frac{g}{\sigma})\right]=N|g|^2r\left(\frac{g}{\sigma}\right),
\end{align}

Furthermore, we have
\begin{align}
{\rm E}_{\psi}^2[\bar{\Delta}]&=g^4{\rm E}_{\psi}^2\left[\left(\sum\limits_{n=0}^{N-1}\chi(n,\omega,\psi,\frac{g}{\sigma})n\right)\right]=g^4\left(\sum\limits_{n=0}^{N-1}n\right)^2r^2\left(\frac{g}{\sigma}\right).
\end{align}

Thus
\begin{align}\label{jessres}
&\frac{2}{\sigma^2}\frac{{\rm E}_{\psi}^2[\bar{\Delta}]}{{\rm E}_{\psi}[\bar{H}]}=\frac{2g^2}{\sigma^2}\frac{\left(\sum\limits_{n=0}^{N-1}n\right)^2}{N}r\left(\frac{g}{\sigma}\right).
\end{align}

According to (\ref{dTd}), (\ref{jessbound}) and (\ref{jessres}), ${\rm ECRB}^{-1}({\omega})$ (\ref{ECRBinv}) can be approximated as
\begin{align}\label{ECRBinv1}
&{\rm ECRB}^{-1}({\omega})=\frac{2T}{\sigma^2}{\rm E}_{\psi}\left[\bar{\mathbf d}^{\rm T}{\boldsymbol\Lambda}\bar{\mathbf d}-\frac{\bar{\Delta}^2}{\bar{H}}\right]\notag\\
&\leq(\approx)\frac{2|g|^2T}{\sigma^2}\left(\sum\limits_{n=0}^{N-1}n^2\right)r\left(\frac{g}{\sigma}\right)-
\frac{2g^2T}{\sigma^2}\frac{\left(\sum\limits_{n=0}^{N-1}n\right)^2}{N}r\left(\frac{g}{\sigma}\right)\notag\\
&=\frac{2|g|^2}{\sigma^2}r\left(\frac{g}{\sigma}\right)\frac{N(N-1)(N+1)}{12}
\end{align}
due to $\sum\limits_{n=0}^{N-1}n^2=\frac{(N-1)N(2N-1)}{6}$ and $\sum\limits_{n=0}^{N-1}n=\frac{N(N-1)}{2}$.

The Chernoff bound \cite{Proakis} $ \Phi(x)\Phi(-x)\leq \frac{1}{4}{\rm e}^{-\frac{x^2}{2}}$ can be used to approximate $h(x;\sigma^2)$ as
\begin{align}\label{chernoffbound}
h(x;\sigma^2)=\frac{1}{2\pi}\frac{{\rm e}^{-\frac{2x^2}{\sigma^2}}}{\Phi\left(\frac{x}{\sigma/\sqrt{2}}\right)\Phi\left(\frac{-x}{\sigma/\sqrt{2}}\right)}\geq (\approx) \frac{2}{\pi}{\rm e}^{-\frac{x^2}{\sigma^2}},
\end{align}
which is also a very tight approximation. Utilizing the integration \cite{MartinTom}
\begin{align}
&\int_{0}^{2\pi}{\rm e}^{-x\cos^2\gamma}\sin^2(\gamma){\rm d}\gamma=\int_{0}^{2\pi}{\rm e}^{-x\sin^2\gamma}\cos^2(\gamma){\rm d}\gamma=\pi{\rm e}^{-\frac{x}{2}}\left(I_0\left(\frac{x}{2}\right)+I_1\left(\frac{x}{2}\right)\right),
\end{align}
where $I_v(\cdot)$ are the modified Bessel functions of the first kind, $r\left(\frac{g}{\sigma}\right)$ can be approximated as
\begin{align}\label{rapp}
&r\left(\frac{g}{\sigma}\right)={\rm E}_{\psi}\left[\chi(n,\omega,\psi,\frac{g}{\sigma})\right]\stackrel{a}\leq (\approx)\frac{2}{\pi}{\rm E}_{\psi}\left[\sin^2(n\omega+\psi){\rm e}^{-\frac{g^2\cos^2(n\omega+\psi)}{\sigma^2}}+\cos^2(n\omega+\psi){\rm e}^{-\frac{g^2\sin^2(n\omega+\psi)}{\sigma^2}}\right]\notag\\
&=\frac{2}{\pi}{\rm e}^{\frac{-g^2}{2\sigma^2}}\left(I_0\left(\frac{g^2}{2\sigma^2}\right)+I_1\left(\frac{g^2}{2\sigma^2}\right)\right),
\end{align}
where $\stackrel{a}\leq$ is due to (\ref{chernoffbound}). Substituting (\ref{rapp}) in (\ref{ECRBinv1}), ${\rm ECRB}^{-1}({\omega})$ can be further approximated as
\begin{align}\label{ELCRBinv}
&{\rm ECRB}^{-1}({\omega})\leq (\approx) \frac{2|g|^2T}{\sigma^2}r\left(\frac{g}{\sigma}\right)\frac{N(N-1)(N+1)}{12}\notag\\
&\leq (\approx)\frac{4}{\pi}\frac{|g|^2T}{\sigma^2}{\rm e}^{\frac{-g^2}{2\sigma^2}}\left(I_0\left(\frac{g^2}{2\sigma^2}\right)+I_1\left(\frac{g^2}{2\sigma^2}\right)\right)\frac{N(N-1)(N+1)}{12}\triangleq {\rm ELCRB}^{-1}({\omega}).
\end{align}
Therefore
\begin{align}
{\rm ECRB}({\omega})\geq(\approx) {\rm ELCRB}({\omega}),
\end{align}
i.e., ${\rm ELCRB}({\omega})$ is a lower bound of ${\rm ECRB}({\omega})$ and approximates ${\rm ECRB}({\omega})$ well.

For low SNR scenario, $I_0(x)\approx 1$ and $I_1(x)\approx 0$, ${\rm ELCRB}^{-1}({\omega})$ (\ref{ELCRBinv}) is approximated as
\begin{align}\label{ELCRBlowSNR}
{\rm ELCRB}^{-1}({\omega})\approx \frac{4}{\pi}\frac{|g|^2}{\sigma^2}\frac{N(N-1)(N+1)}{12}
\end{align}
and
\begin{align}
{\rm ELCRB}({\omega})=\frac{\pi}{2}{\rm CRB}({\omega}).
\end{align}
For high SNR scenario, one has
\begin{align}\label{Ivapp}
I_v(x)\approx\frac{{\rm e}^x}{\sqrt{2\pi x}}.
\end{align}
Substituting (\ref{Ivapp}) into (\ref{ELCRBinv}) yields
\begin{align}\label{ELCRBhighSNR}
{\rm ELCRB}^{-1}({\omega})&\approx \frac{8}{\pi\sqrt{\pi}}\frac{g}{\sigma}\frac{N(N-1)(N+1)}{12}
\approx \frac{2}{3\pi^{\frac{3}{2}}}\frac{g}{\sigma}N^3.
\end{align}
According to (\ref{ELCRBlowSNR}) and (\ref{ELCRBhighSNR}), we have
\begin{align}\label{asymELCRB}
{\rm ELCRB}({\omega})\approx\begin{cases}
\frac{3\pi}{{\rm SNR}N^3T},{\rm low~SNR},\\
\frac{3\pi^{\frac{3}{2}}}{2}\frac{1}{\sqrt{\rm SNR}N^3T},{\rm high~SNR}.
\end{cases}
\end{align}
Because ${\rm ELCRB}({\omega})$ approximates ${\rm CRB}({\omega})$ well, it can be concluded that under one-bit quantization, the CRB is inversely proportional to the number of snapshots and the cubic of the number of antennas. For lower SNR scenario, the CRB is inversely proportional to SNR, while the CRB is inversely proportional to the square root of the SNR for high SNR scenario.
\section{MVALSE-EP Algorithm}\label{Algorithm}
In this section, MVALSE-EP algorithm is developed to estimate the frequencies. The key step in designing the MVALSE-EP is to introduce the suitable hidden variable and expand the factor graph. Here we define
\begin{align}
{\mathbf z}(t)\triangleq {\mathbf A}({\boldsymbol \omega}){\mathbf x}(t),
\end{align}
and the EP is adopted to iteratively approximate the nonlinear measurement model as a pseudo linear measurement model. In addition, EM and VALSE are adopted to estimate the nuisance parameters and the frequencies. In the following text, we first present the modeling setup. Then we adopt the modularized point of view to design MVALSE-EP algorithm based on existing superresolution VALSE approach.
\subsection{Modeling Setup}
Since the number of targets $K$ is usually unknown, an overcomplete model in which the signal consisting of $N$  targets is assumed \cite{VALSE}
\begin{align}\label{signal-model}
{\mathbf z}(t)=\sum\limits_{k=1}^N {\mathbf a}({\omega}_k)x_k(t)= {\mathbf A}({\boldsymbol \omega}){\mathbf x}(t),
\end{align}
where ${\mathbf A}({\boldsymbol \omega})=[{\mathbf a}({\omega}_1),\cdots,{\mathbf a}({\omega}_N)]$ and satisfies $N> K$. To model the unknown nature of $K$, the binary hidden variables ${\mathbf s}(t)$ are introduced, where $s_{k}(t)=1$ means that the $k$th target is active in the $t$th snapshot, otherwise deactive (${x}_{k}(t)=0$). In addition, it is required that the number of sources are the same across the snapshots, which requires that
\begin{align}
\sum\limits_{k=1}^Ns_{k}(t)=K.
\end{align}
The probability mass function of $s_{k}(t)$ is
\begin{align}\label{sprob}
p(s_{k}(t)) = \rho^{s_{k}(t)}(1-\rho)^{(1-s_{k}(t))},\quad s_{k}(t)\in\{0,1\}.
\end{align}
Given that $s_{k}(t)=1$, we assume that ${x}_{k}(t)\sim {\mathcal {CN}}({x}_{k}(t);0,\tau)$. Thus $(s_{k}(t),{x}_{k}(t))$ follows a Bernoulli-Gaussian distribution, that is
\begin{align}
p({x}_{k}(t)|s_{k}(t);\tau) = (1 - s_{k}(t)){\delta}({x}_{k}(t)) + s_{k}(t){\mathcal {CN}}({x}_{k}(t);0,\tau).\label{pdfw}
\end{align}
From (\ref{sprob}) and (\ref{pdfw}), it can be seen that the parameter $\rho$ denotes the probability of the $k$th component being active and $\tau$ is a variance parameter. The variable ${\boldsymbol \omega} = [\omega_1,...,\omega_N]^{\rm T}$ has the prior PDF $p({\boldsymbol \omega}) = \begin{matrix} \prod_{k=1}^N p(\omega_k) \end{matrix}$. Without any knowledge of the frequency $\omega$, the uninformative prior distribution $p(\omega_k) = {1}/({2\pi})$ is used \cite{VALSE}. For encoding the prior distribution, please refer to \cite{VALSE, MVALSE} for further details. Note that the modeling setup of (\ref{pdfw}) is different from that of \cite{MVALSE}. In \cite{MVALSE}, only binary variables $\{s_{k}\}_{k=1}^N$ are introduced, and such modeling makes the algorithm overestimate the model order.

Given $\mathbf Z$, the PDF $p({\mathbf Y}|{\mathbf Z};\sigma_w^2)=\prod\limits_{t=1}^Tp({\mathbf y}(t)|{\mathbf z}(t);\sigma_w^2)=\prod\limits_{m=1}^N\prod\limits_{t=1}^Tp(y_{m}(t)|z_{m}(t);\sigma_w^2)$ of $\mathbf Y$ can be easily calculated through (\ref{quantmodel}). Let
\begin{align}
&{\boldsymbol \Omega}=(\omega_1,\dots,\omega_N,({\mathbf X},{\mathbf s})),\\
&{\boldsymbol \beta} = \{\sigma_w^2, \rho,~\tau, K\}
\end{align}
be the set of all random variables and the model parameters, respectively. According to the Bayes rule, the joint PDF $p({\mathbf Y },{\mathbf Z},{\boldsymbol \Omega};{\boldsymbol \beta})$ is
\begin{align}\label{jointpdf}
&p({\mathbf Y },{\mathbf Z},{\boldsymbol \Omega};{\boldsymbol \beta})=p({\mathbf Y}|{\mathbf Z})\delta({\mathbf Z}-{\mathbf A}({\boldsymbol \theta}){\mathbf X})\prod\limits_{k=1}^N p(\omega_k\prod\limits_{t=1}^Tp({x}_k(t)|s_k(t))p(s_k(t))\delta\left(\sum\limits_{k=1}^Ns_{k}(t)-K\right).
\end{align}
Given the above joint PDF (\ref{jointpdf}), the type II maximum likelihood (ML) estimation of the model parameters $\hat{\boldsymbol\beta}_{\rm ML}$ is
\begin{align}\label{MLbeta}
\hat{\boldsymbol\beta}_{\rm ML}=\underset{\boldsymbol \beta}{\operatorname{argmax}}~ p({\mathbf Y };{\boldsymbol \beta})=\underset{\boldsymbol \beta}{\operatorname{argmax}}~\int p({\mathbf Y },{\mathbf Z},{\boldsymbol \Omega};{\boldsymbol \beta}){\rm d}{\mathbf Z}{\rm d}{{\boldsymbol \Omega}}.
\end{align}
Then the minimum mean squared error (MMSE) estimates of the parameters $({\mathbf Z},{\boldsymbol \Omega})$ is
\begin{align}\label{MMSE}
(\hat{\mathbf Z},\hat{{\boldsymbol \Omega}})={\rm E}[({\mathbf Z},{\boldsymbol \Omega})| {\mathbf Y};{\boldsymbol \beta}_{\rm ML}],
\end{align}
 where the expectation is taken with respect to
\begin{align}
p({\mathbf Z},{\boldsymbol \Omega}|{\mathbf Y };\hat{\boldsymbol \beta}_{\rm ML})=\frac{p({\mathbf Z},{\boldsymbol \Omega},{\mathbf Y };\hat{\boldsymbol \beta}_{\rm ML})}{p({\mathbf Y };\hat{\boldsymbol \beta}_{\rm ML})}
\end{align}
Directly solving the ML estimate of $\boldsymbol \beta$ (\ref{MLbeta}) or the MMSE estimate of $({\mathbf Z},{\boldsymbol \Omega})$ (\ref{MMSE}) are both intractable. As a result, an iterative algorithm is designed in the ensuing text.

\subsection{Algorithm Design}
\begin{figure*}
\centering
\includegraphics[width=5.0in]{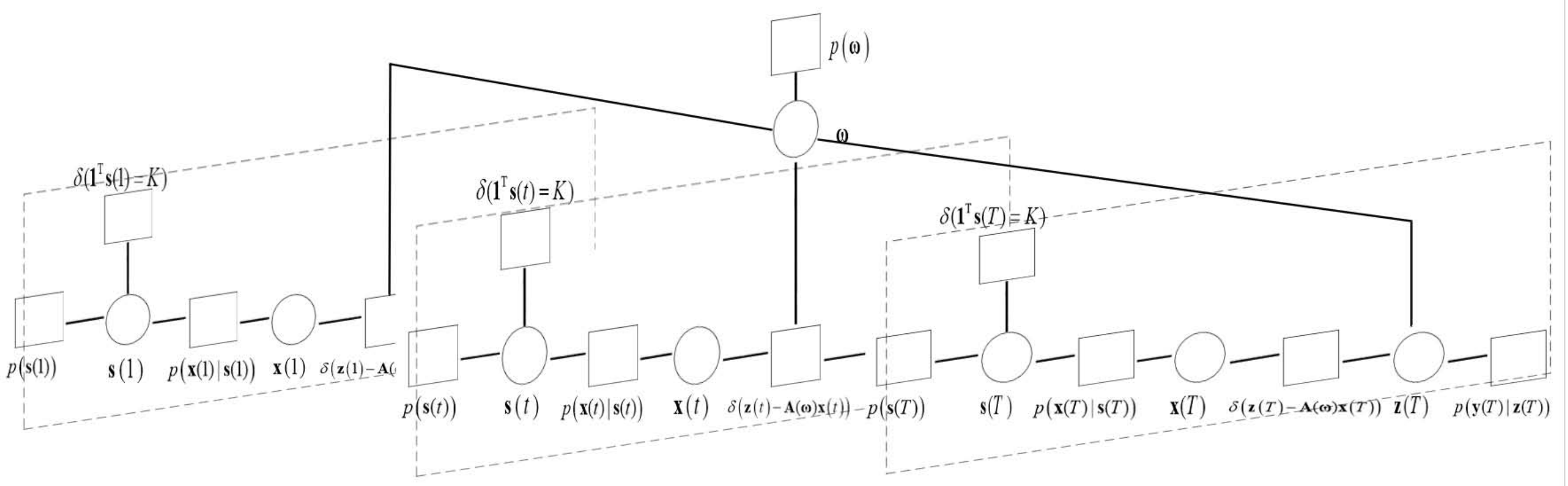}
\caption{The factor graph. The factor graph for each snapshot is almost the same and share the same variable node $\boldsymbol\omega$ and the factor node $p(\boldsymbol\omega)$.}
\label{FC_fig}                                                                                                                                                                                                                                                       \end{figure*}

\begin{figure}[h!t]
\centering
\includegraphics[width=3.6in]{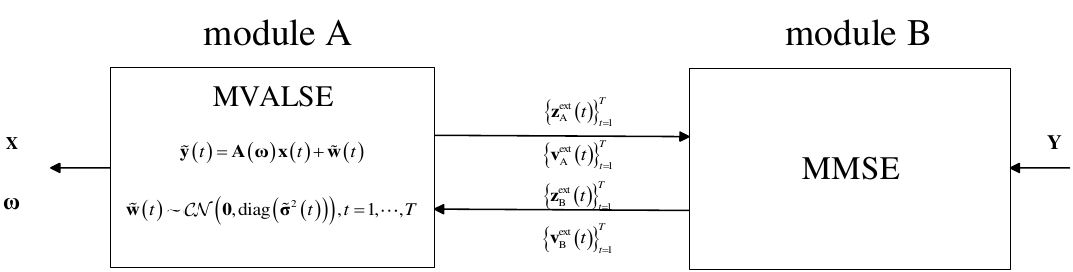}
\caption{The module of the MVALSE-EP algorithm. From the modularized point of view, the algorithm can be decomposed as two modules named module A and module B, where module A corresponds to the standard linear model, and module B corresponds to the MMSE estimation. By iterating between the two modules, the frequencies and amplitudes are refined gradually.}
\label{modulepic}                                                                                                                                                                                                                                                       \end{figure}

Below we describe the details of the algorithm and show that how to utilize the basic VALSE algorithm to perform the DOAs estimation. For brevity, the iteration index is omitted. The factor graph of the joint PDF (\ref{jointpdf}) is shown in Fig. \ref{FC_fig}, and the algorithm is designed according to Fig. \ref{modulepic}.
Firstly, initialize the message transmitted from the factor node $\delta({\mathbf Z}-{\mathbf A}{\mathbf X})$ to the variable node ${\mathbf Z}$ as
\begin{align}
&{m}_{\delta\rightarrow {\mathbf Z}}({\mathbf Z})=\prod\limits_{t=1}^T{m}_{\delta\rightarrow {\mathbf z}(t)}({\mathbf z}(t))=\prod\limits_{t=1}^T{\mathcal{CN}}({\mathbf z}(t);{\mathbf z}_{{\rm A}}^{\rm ext}(t),{\rm diag}({\mathbf v}_{{\rm A}}^{\rm ext}(t))).
\end{align}

\subsubsection{Componentwise MMSE}
According to EP, the message $m_{{\mathbf Z}\rightarrow \delta}({\mathbf Z})$ transmitted from the variable node ${\mathbf Z}$ to the factor node $\delta({\mathbf Z}-{\mathbf A}{\mathbf X})$ can be calculated as \cite{Minka}
\begin{subequations}\label{extBz0}
\begin{align}
m_{{\mathbf Z}\rightarrow \delta}({\mathbf Z})&\propto \frac{{\rm Proj}[{m}_{\delta\rightarrow {\mathbf Z}}({\mathbf Z})p({\mathbf Y}|{\mathbf Z})]}{{m}_{\delta\rightarrow {\mathbf Z}}({\mathbf Z})}\propto \frac{\prod\limits_{t=1}^T{\rm Proj}[{m}_{\delta\rightarrow {\mathbf z}(t)}({\mathbf z}(t))p({\mathbf y}(t)|{\mathbf z}(t))]}{\prod\limits_{t=1}^T{m}_{\delta\rightarrow {\mathbf z}(t)}({\mathbf z}(t))}\notag\\
&\triangleq \prod\limits_{t=1}^T\frac{{\rm Proj}[q_{{\rm B}}({\mathbf z}(t))]}{{m}_{\delta\rightarrow {\mathbf z}(t)}({\mathbf z}(t))} \triangleq \prod\limits_{t=1}^T m_{{\mathbf z}(t)\rightarrow \delta}({\mathbf z}(t)),
\end{align}
\end{subequations}
where $\propto $ denotes identity up to a normalizing constant. First, the MMSE estimate of ${\mathbf z}(t)$ can be obtained, i.e.,
\begin{align}
&{\mathbf z}_{{\rm B}}^{\rm post}(t)={\rm E}[{\mathbf z}(t)|q_{{\rm B}}({\mathbf z}(t))],\label{comb_means}\\
&{\mathbf v}_{{\rm B}}^{\rm post}(t)={\rm Var}[{\mathbf z}(t)|q_{{\rm B}}({\mathbf z}(t))]\label{comb_vars},
\end{align}
where ${\rm E}[\cdot|q_{{\rm B}}({\mathbf z}(t))]$ and ${\rm Var}[\cdot|q_{{\rm B}}({\mathbf z}(t))]$ are the mean and variance operations taken componentwise with respect to the distribution $\propto q_{{\rm B}}({\mathbf z}(t))$. Here we adopt the diagonal EP and ${\rm Proj}[q_{{\rm B}}({\mathbf z}(t))]$ is
\begin{align}\label{postBz}
{\rm Proj}[q_{{\rm B}}({\mathbf z}(t))]={\mathcal {CN}}({\mathbf z}(t);{\mathbf z}_{{\rm B}}^{\rm post}(t),{\rm diag}({\mathbf v}_{{\rm B}}^{\rm post}(t))).
\end{align}
Substituting (\ref{postBz}) in (\ref{extBz0}), the message $m_{{\mathbf z}\rightarrow \delta}({\mathbf z}(t))$ from the variable node ${\mathbf z}(t)$ to the factor node $\delta({\mathbf z}-{\mathbf A}({\boldsymbol \omega}){\mathbf x}(t))$ is calculated as
\begin{align}\label{extBtoA}
m_{{\mathbf z}(t)\rightarrow \delta}({\mathbf z}(t))&\propto \frac{{\mathcal {CN}}({\mathbf z}(t);{\mathbf z}_{{\rm B}}^{\rm post}(t),{\rm diag( {\mathbf v}_{{\rm B}}^{\rm post}(t))})}{{\mathcal {CN}}({\mathbf z}(t);{\mathbf z}_{{\rm A}}^{\rm ext}(t),{\rm diag ({\mathbf v}_{{\rm A}}^{\rm ext}(t))})}\propto {\mathcal {CN}}({\mathbf z}(t);{\mathbf z}_{{\rm B}}^{\rm ext}(t),{\rm diag }({\mathbf v}_{{\rm B},{\mathbf z}(t)}^{\rm ext}(t))),
\end{align}
where ${\mathbf z}_{{\rm B}}^{\rm ext}(t)$ and ${\mathbf v}_{{\rm B}}^{\rm ext}(t)$ are \cite{UnifiedSPL}
\begin{subequations}
\begin{align}
&{\mathbf v}_{{\rm B}}^{\rm ext}(t)=\left(\frac{1}{{\mathbf v}_{{\rm B},{\mathbf z}(t)}^{\rm post}(t)}-\frac{1}{{\mathbf v}_{{\rm A},{\mathbf z}(t)}^{\rm ext}(t)}\right)^{-1},\label{extB_var}\\
&{\mathbf z}_{{\rm B}}^{\rm ext}(t)={\mathbf v}_{{\rm B}}^{\rm ext}(t)\odot\left(\frac{{\mathbf z}_{{\rm B}}^{\rm post}(t)}{{\mathbf v}_{{\rm B}}^{\rm post}(t)}-\frac{{\mathbf z}_{{\rm A}}^{\rm ext}(t)}{{\mathbf v}_{{\rm A}}^{\rm ext}(t)}\right),\label{extB_mean}
\end{align}
\end{subequations}
where $\odot$ denotes componentwise multiplication. Consequently, we have
\begin{align}\label{extBtoAall}
&m_{{\mathbf Z}\rightarrow \delta}({\mathbf Z})\propto \prod\limits_{t=1}^Tm_{{\mathbf z}(t)\rightarrow \delta}({\mathbf z}(t))\propto p\prod\limits_{t=1}^T{\mathcal {CN}}({\mathbf z}(t);{\mathbf z}_{{\rm B}}^{\rm ext}(t),{\rm diag }({\mathbf v}_{{\rm B}}^{\rm ext}(t))).
\end{align}

\subsubsection{MVALSE module}
According to (\ref{extBtoAall}), the message $m_{{\mathbf Z}\rightarrow \delta}({\mathbf Z})$ transmitted from the variable node ${\mathbf Z}$ to the factor node $\delta({\mathbf Z}-{\mathbf A}{\mathbf X})$ is Gaussian distributed and is independent of the snapshot $t$. Based on the definition of the factor node $\delta({\mathbf Z}-{\mathbf A}{\mathbf X})$, $T$ pseudo linear observation models
\begin{align}\label{heomodel}
\tilde{\mathbf y}(t)={\mathbf A}({\boldsymbol \omega}){\mathbf x}(t)+\tilde{\mathbf w}(t),\quad t=1,\cdots,T
\end{align}
are obtained, where $\tilde{\mathbf w}(t)\sim {\mathcal {CN}}({\mathbf 0},{\rm diag}(\tilde{\boldsymbol \sigma}^2(t)))$, $\tilde{\mathbf y}(t)={\mathbf z}_{{\rm B}}^{\rm ext}(t)$ and $\tilde{\boldsymbol \sigma}^2(t)={\mathbf v}_{{\rm B}}^{\rm ext}(t)$. For the $t$th equation in (\ref{heomodel}), the variances of the heteroscedastic noise $\tilde{\mathbf w}(t)$ are different. In addition, All the snapshots share the same frequency ${\boldsymbol \omega}$. Then we run the MVALSE under heteroscedastic noise algorithm \cite{QiJOE} with known noise variances. Note that we could average the noise variance with respect to snapshots or measurements. Such operation simplifies the computation especially when averaging over snapshots. The only difference is that for this modeling setup, the $k$th target is active only when
\begin{align}\label{delta-k-active}
\Delta_k = \frac{1}{T}\sum_{t=1}^T\left(\ln\frac{v_{k}(t)}{\tau} + \frac{|u_{k}(t)|^2}{v_{k}(t)}\right)+\ln\frac{\rho}{1-\rho}>0,
\end{align}
where $v_{k}(t)$ and $u_{k}(t)$ are
\begin{align}\label{v-k-u-k}
&v_{k}(t) = \left({\rm tr}(\boldsymbol\Sigma^{-1}(t))+\frac{1}{\tau}-{\mathbf J}_{{\mathcal S},k}^{\rm H}(t)({\mathbf J}_{{\mathcal S}}^{\rm H}+\frac{1}{\tau}\mathbf I_{|\mathcal S|})^{-1}{\mathbf J}_{{\mathcal S},k}^{\rm H}\right)^{-1},\notag\\
&u_{k}(t) = v_{k}(t)\left({h}_{k}(t)-{\mathbf J}_{{\mathcal S},k}^{\rm H}(t)({\mathbf J}_{{\mathcal S}}(t)+\frac{1}{\tau}\mathbf I_{|\mathcal S|})^{-1}{\mathbf h}_{\mathcal S}(t)\right),
\end{align}
$\boldsymbol\Sigma^{-1}(t)={\rm diag}\left(\tilde{\boldsymbol \sigma}^2(t)\right)$, ${\mathbf J}(t)$ and ${\mathbf h}(t)$ are
\begin{subequations}\label{J-H}
\begin{align}
&{J}_{i,j}(t)= 	
\begin{cases}
{\rm tr}({\boldsymbol\Sigma}^{-1}(t)),&i=j\\
{\widehat{\mathbf a}}^{\rm H}_i{\boldsymbol\Sigma}^{-1}(t){\widehat{\mathbf a}}_j,&i\neq{j}
\end{cases},\quad i,j\in\{1,\cdots,N\},\\
&{\mathbf h}(t) = \widehat{\mathbf A}^{\rm H}{\boldsymbol\Sigma}^{-1}(t)\tilde{\mathbf y}(t),
\end{align}
\end{subequations}
where ${J}_{i,j}(t)$ denotes the $(i,j)$th element of ${\mathbf J}(t)$ and ${\widehat{\mathbf a}}_j$ denotes the posterior mean of ${\mathbf a}_j$. The deactive case can be obtained in a similar way. Once the posterior PDF $q({\boldsymbol \omega}|\tilde{\mathbf Y})$, $q({\mathbf X}|\tilde{\mathbf Y})$ and the model parameters $\tau$ and $\rho$ are obtained, the posterior means and variances of ${\mathbf z}(t)={\mathbf A}({\boldsymbol \omega}){\mathbf x}(t)$ are also obtained as \cite{VALSEEP}
\begin{align}
&{\mathbf z}_{{\rm A}}^{\rm post}(t)=\hat{\mathbf A}_{\hat{S}}\hat{\mathbf x}_{\hat{S}}(t),\label{post_means_A}\\
&{\mathbf v}_{{\rm A}}^{\rm post}(t)= {\rm diag}(\hat{\mathbf A}_{\hat{S}}\hat{\mathbf C}_{\hat{S}}(t)\hat{\mathbf A}_{\hat{S}}^{\rm H})+\left(\hat{\mathbf x}_{\hat{S}}^{\rm H}(t)\hat{\mathbf x}_{\hat{S}}(t){\mathbf 1}_{N}-|\hat{\mathbf A}_{\hat{S}}|^2|\hat{\mathbf x}_{\hat{S}}(t)|^2\right)p+{\rm tr}(\hat{\mathbf C}_{\hat{S}}(t)){\mathbf 1}_N-|\hat{\mathbf A}_{\hat{S}}|^2{\rm diag}(\hat{\mathbf C}_{\hat{S}}(t)),\label{post_vars_A}
\end{align}
where $\hat{\mathbf x}_{\hat{S}}(t)$ and $\hat{\mathbf C}_{\hat{S}}(t)$ are the posterior means and covariance matrix of ${\mathbf x}(t)$, $\hat{\mathbf A}_{\hat{S}}$ is the estimate of ${\mathbf A}_{\hat{S}}$. The noise variance $\sigma^2$ can be obtained via the EM algorithm as
\begin{align}\label{noisevarest}
\sigma^2=\frac{1}{NT}\sum\limits_{t=1}^T\left({\|\tilde{\mathbf y}(t)-{\mathbf z}_{{\rm A}}^{\rm post}(t)\|^2+{\mathbf 1}^{\rm T}{\mathbf v}_{{\rm A}}^{\rm post}(t)}\right).
\end{align}
Then we calculate the message ${m}_{\delta\rightarrow {\mathbf z}(t)}({\mathbf z}(t))$ as
\begin{align}\label{extA}
{m}_{\delta\rightarrow {\mathbf z}(t)}({\mathbf z}(t))= \frac{{\rm Proj}[q_{\rm A}({\mathbf z}(t))]}{m_{{\mathbf z}(t)\rightarrow \delta}({\mathbf z}(t))}.
\end{align}
where ${\rm Proj}[q_{\rm A}({\mathbf z}(t))]$ is
\begin{align}
{\rm Proj}[q_{{\rm A}}({\mathbf z}(t))]={\mathcal {CN}}({\mathbf z}(t);{\mathbf z}_{{\rm A}}^{\rm post}(t),{\rm diag}({\mathbf v}_{{\rm A}}^{\rm post}(t))).
\end{align}
According to (\ref{extA}), ${m}_{\delta\rightarrow {\mathbf z}(t)}({\mathbf z}(t))$ is calculated to be
\begin{align}
{m}_{\delta\rightarrow {\mathbf z}(t)}({\mathbf z}(t))={\mathcal {CN}}({\mathbf z}(t);{\mathbf z}_{{\rm A}}^{\rm ext}(t),{\rm diag}({\mathbf v}_{{\rm A}}^{\rm ext}(t))),
\end{align}
where the extrinsic ${\mathbf z}_{{\rm A}}^{\rm ext}(t)$ and variance ${\mathbf v}_{{\rm A}}^{\rm ext}(t)$ are given by \cite{UnifiedSPL}
\begin{align}
&\frac{1}{{\mathbf v}_{{\rm A}}^{\rm ext}(t)}= \frac{1}{{\mathbf v}_{{\rm A}}^{\rm post}(t)}-\frac{1}{\tilde{\boldsymbol \sigma}^2(t)},\label{extvarA}\\
&{\mathbf z}_{{\rm A}}^{\rm ext}(t)={{\mathbf v}_{{\rm A}}^{\rm ext}(t)}\odot\left(\frac{{\mathbf z}_{{\rm A}}^{\rm post}(t)}{{\mathbf v}_{{\rm A}}^{\rm post}(t)}-\frac{\tilde{\mathbf y}(t)}{\tilde{\boldsymbol \sigma}^2(t)}\right),\label{extmeanA}
\end{align}
and we input them to module B. The algorithm is closed and the algorithm iterates until convergence or the maximum number of iterations is reached. The MVALSE-EP algorithm is summarized as Algorithm \ref{MVALSE-EP}. For further details, please refer to \cite{VALSEEP}.
\begin{algorithm}[h]
\caption{MVALSE-EP algorithm}\label{MVALSE-EP}
\begin{algorithmic}[1]
\STATE Initialize ${\mathbf v}_{{\rm A}}^{\rm ext}(t)={\mathbf 10}^{4}$, ${\mathbf z}_{{\rm A}}^{\rm ext}(t)={\mathbf 0}_M$, $t=1,\cdots,T$;
\STATE Initialize the noise variance $\sigma^2$.
\STATE Perform the MMSE estimate of $\mathbf Z$ in module B and calculate the extrinsic massage from module B to module A.
\STATE Initialize $\rho$, $\hat{K}$ and $\tau$.
\STATE Initialize $q({\omega}_i|\tilde{\mathbf Y})$ and obtain ${\mathbf J}(t)$, ${\mathbf h}(t)$.
\STATE Set the number of outer iterations ${\rm Iter}_{\rm max}$;  \
\FOR {${\rm Iter}=1,\cdots,{\rm Iter}_{\rm max}$ }
\STATE Update the support $\mathbf s$.
\STATE Update $\tau$ and $\rho$.
\STATE Refine the mean and concentration parameters of the frequencies $\omega_i$, $i\in {\mathcal S}$.
\STATE Calculate the posterior means ${\mathbf z}_{{\rm A}}^{\rm post}(t)$ (\ref{post_means_A}) and variances ${\mathbf v}_{{\rm A}}^{\rm post}(t)$ (\ref{post_vars_A}).
\STATE Compute the extrinsic mean and variance of ${\mathbf z}(t)$ as ${\mathbf v}_{{\rm A}}^{\rm ext}(t)$ (\ref{extvarA}), ${\mathbf z}_{{\rm A}}^{\rm ext}(t)$ (\ref{extmeanA}).\
\STATE Compute the post mean and variance of ${\mathbf z}(t)$ as ${\mathbf z}_{{\rm B}}^{\rm post}(t)$ (\ref{comb_means}), ${\mathbf v}_{{\rm B}}^{\rm post}(t)$ (\ref{comb_vars}), $t=1,\cdots,T$.
\STATE Compute the extrinsic mean and variance of ${\mathbf z}(t)$ as ${\mathbf z}_{{\rm B}}^{\rm ext}(t)$ (\ref{extB_mean}) and ${\mathbf v}_{{\rm B}}^{\rm ext}(t)$ (\ref{extB_var}), and set $\tilde{\boldsymbol \sigma}^2(t) ={\mathbf v}_{{\rm B}}^{\rm ext}(t)$ and $\tilde{\mathbf y}(t)={\mathbf z}_{{\rm B}}^{\rm ext}(t)$.
\STATE Estimate the noise variance $\sigma^2$ (\ref{noisevarest}).
\STATE Update ${\mathbf J}(t)$ and ${\mathbf h}(t)$.
\ENDFOR
\STATE Return $\hat{\boldsymbol \omega}$, $\hat{\mathbf X}$, $\hat{\mathbf Z}$ and $\hat{K}$.
\end{algorithmic}
\end{algorithm}
\subsection{Computation Complexity}

Let ${\rm Iter}_{\rm max}$ denote the number of iterations. For the proposed MVALSE-EP algorithm shown in Algorithm \ref{MVALSE-EP}, it consists of main steps: The initialization, the componentwise MMSE operation, the mulisnapshot VALSE (MVALSE) algorithm. For initialization, the computation complexity is $O(N^2T)$ to obtain the posterior PDFs of the frequencies. The complexity of the MMSE module per iteration is $O(NT)$. As for the MVALSE algorithm, its complexity per iteration is dominated by calculating $\Delta_k$ (\ref{delta-k-active}), whose complexity is $O(N\hat{K}^3T)$ per iteration. Thus the overall computation complexity of the MVALSE-EP is $O((N^2+N\hat{K}^3\times {\rm Iter}_{\rm max})T)$.
\subsection{Further Discussion}
It is worth noting that the $k$th target is active only when (\ref{delta-k-active}) is satisfied. This is different from the MVALSE algorithm \cite{MVALSE} where the $k$th target is active only when
\begin{align}\label{mvalse-delta-k-active}
\Delta_k^{'} = \frac{1}{T}\sum_{t=1}^T\left(\ln\frac{v_{k}(t)}{\tau} + \frac{|u_{k}(t)|^2}{v_{k}(t)}\right)+\frac{1}{T}\ln\frac{\rho}{1-\rho}>0.
\end{align}
Because $\ln\frac{\rho}{1-\rho}<0$ as $\rho<0.5$ in general, it can be seen that with all the parameters being the same, the $k$th target is more likely to be active for the previous MVALSE algorithm \cite{MVALSE}. We have found that if (\ref{mvalse-delta-k-active}) is adopted as the criterion for activating the frequency, it is more likely to generate false alarms corresponding to harmonic components. Thus, the proposed criterion (\ref{delta-k-active}) reduces the false alarms.

It is also numerically found that MVALSE-EP tends to fit some spurious components and overestimates the number of sources for real data. As a consequence, we make the active criterion more harsh when we activate the $k$th frequency by minusing a positive bias $\lambda$ as
\begin{align}\label{delta-k-active_tune}
\frac{1}{T}\sum_{t=1}^T\left(\ln\frac{v_{k}(t)}{\tau} + \frac{|u_{k}(t)|^2}{v_{k}(t)}\right)+\ln\frac{\rho}{1-\rho}-\lambda>0,
\end{align}
where $\lambda>0$ is a tune parameter. For example, $\lambda=5$. Note that increasing $\lambda$ will make the model more sparse.
\section{Numerical Simulation}\label{NS}
In this section, numerical experiments are conducted to evaluate the effectiveness of the proposed algorithm, through comparing with the CRBs. For the numerical simulation part, we conduct the DOA estimation and the element spacing of the linear array is half wavelength. Thus the DOA ${\boldsymbol \theta}$ is related to the frequency $\boldsymbol \omega$ via ${\boldsymbol \omega}=\pi\sin\left({\boldsymbol \theta}\right)$. For the real data case, we evaluate the range estimation for mmWave FMCW system. We evaluate the signal ${\mathbf Z}$ estimation error, the DOA estimation error, the correct model order estimation probability under quantized measurements in numerical simulations.

The phases of the weight coefficients are drawn i.i.d. from a uniform distribution between $[0,2\pi)$. For multi-bit quantization, a uniform quantizer is adopted and the quantization interval is restricted to $[-3\sigma_z,3\sigma_z]$, where $\sigma_z^2$ is the variance of the signal $\Re\{{\mathbf z}_N\}$ or $\Im\{{\mathbf z}_N\}$. In our setting, it can be calculated that $\sigma_z^2\approx \sum\limits_{k=1}^Kg_k^2+\sigma^2$.
For one-bit quantization, zero is chosen as the threshold.
For each target, its SNR (dB) is uniformly drawn from $[{\rm SNR}_{\rm min},{\rm SNR}_{\rm min}+\Delta]$ (dB). The amplitude of the target is determined from the SNRs and is fixed across the snapshots, the phase is uniformly drawn from $[0,2\pi)$. The noninformative prior, i.e., $p(\theta_i)=1/(2\pi)$ is adopted for the MVALSE and MVALSE-EP algorithms. The number of maximum iterations is set as ${\rm Iter}_{\rm max}=50$. The normalized MSE (NMSE) of signal $\hat{\mathbf Z}$ (for unquantized and multi-bit quantized system) and the root MSE (RMSE) of the DOAs $\widehat{\boldsymbol\theta}$ are defined as ${{\rm NMSE}}(\hat{\mathbf Z})\triangleq \|\hat{\mathbf Z} - {\mathbf Z}\|_{\rm F}^2/\|{\mathbf Z}\|_{\rm F}^2$ and ${{\rm RMSE}}(\boldsymbol\theta)\triangleq \sqrt{\|\widehat{\boldsymbol\theta} - \boldsymbol\theta\|_2^2/K}$, respectively. Please note that, due to magnitude ambiguity, it is impossible to recover the exact magnitude of ${x}_k(t)$ from one-bit measurements in the noiseless scenario.  Thus for one-bit quantization, the debiased NMSE is used. The debiased NMSE of the signal defined as ${{\rm dNMSE}}({\mathbf Z})\triangleq \underset{\mathbf c}{\operatorname{min}}~{\|{\mathbf Z}^*-{\rm diag}({\mathbf c})\hat{\mathbf Z}\|_{\rm F}^2}/{\|{\mathbf Z}^*\|_{\rm F}^2}$ are calculated. As for the DOA error, we average only the trials in which all those algorithms estimate the correct model order. All the results are averaged over $300$ Monte Carlo (MC) trials unless stated otherwise. The empirical probability of correct model order estimation ${\rm P}(\hat{K}=K)$ is adopted as a performance metric. The RMSE of the DOA is calculated only when the model order is correctly estimated, i.e., $\hat{K}=K$.
\subsection{Asymptotic Property of CRB for a Single Source}
\begin{figure*}
  \centering
  \subfigure[]{
    \label{CRBvsSNR} 
    \includegraphics[width=52mm]{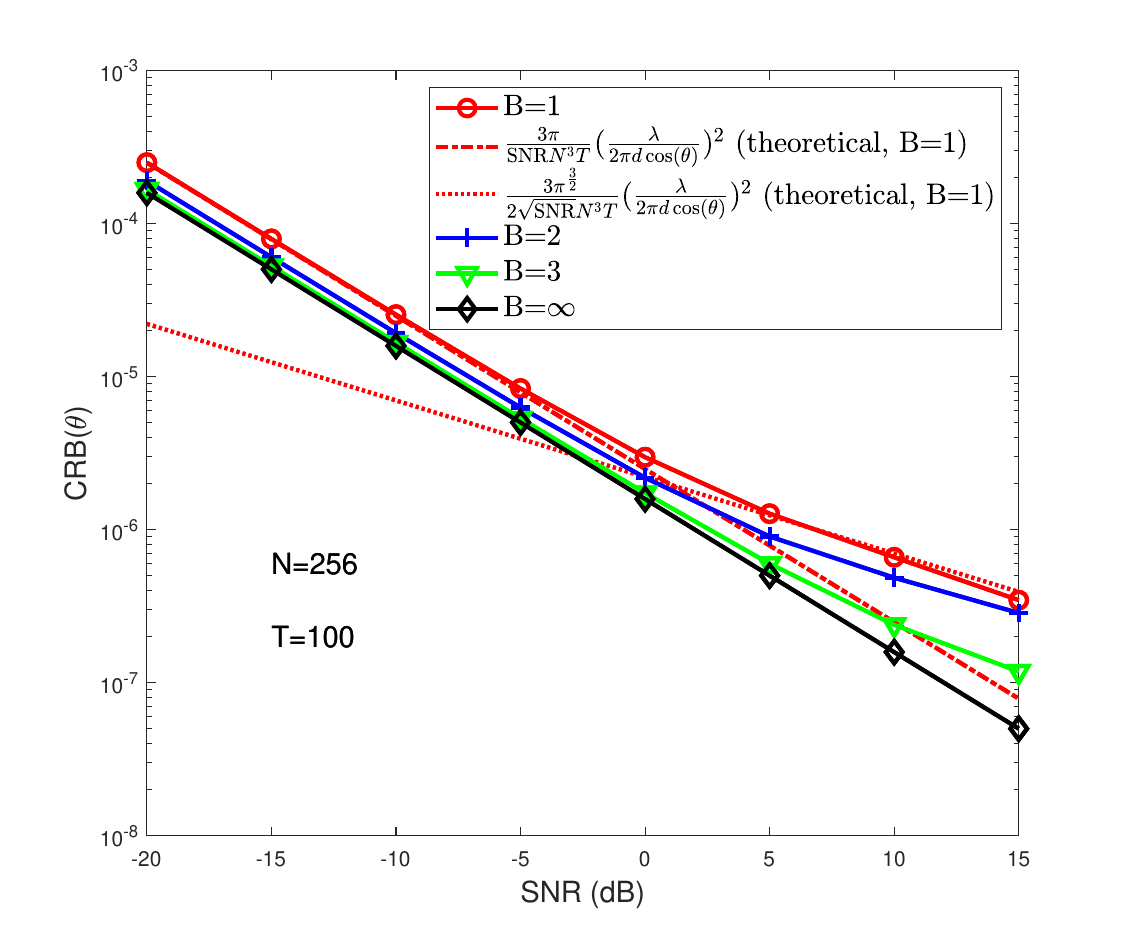}}
  \subfigure[]{
    \label{CRBvsTlowSNR} 
    \includegraphics[width=52mm]{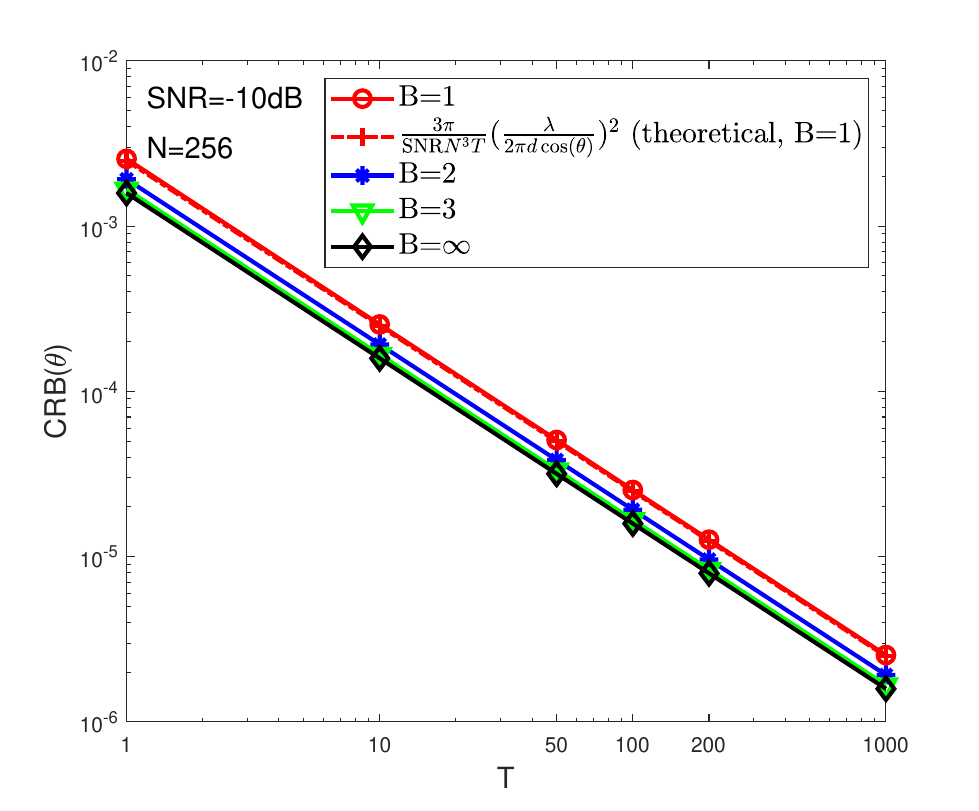}}
    \subfigure[]{
    \label{CRBvsThighSNR} 
    \includegraphics[width=52mm]{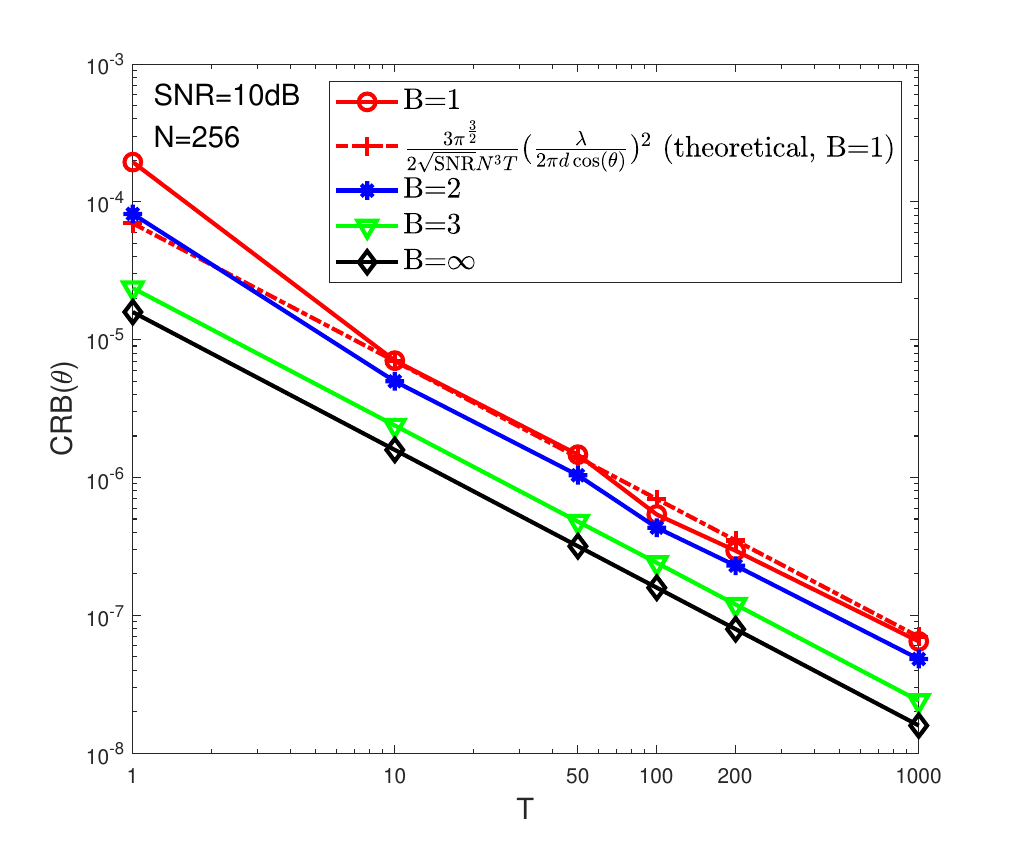}}
  \subfigure[]{
    \label{CRBvsNlowSNR} 
    \includegraphics[width=52mm]{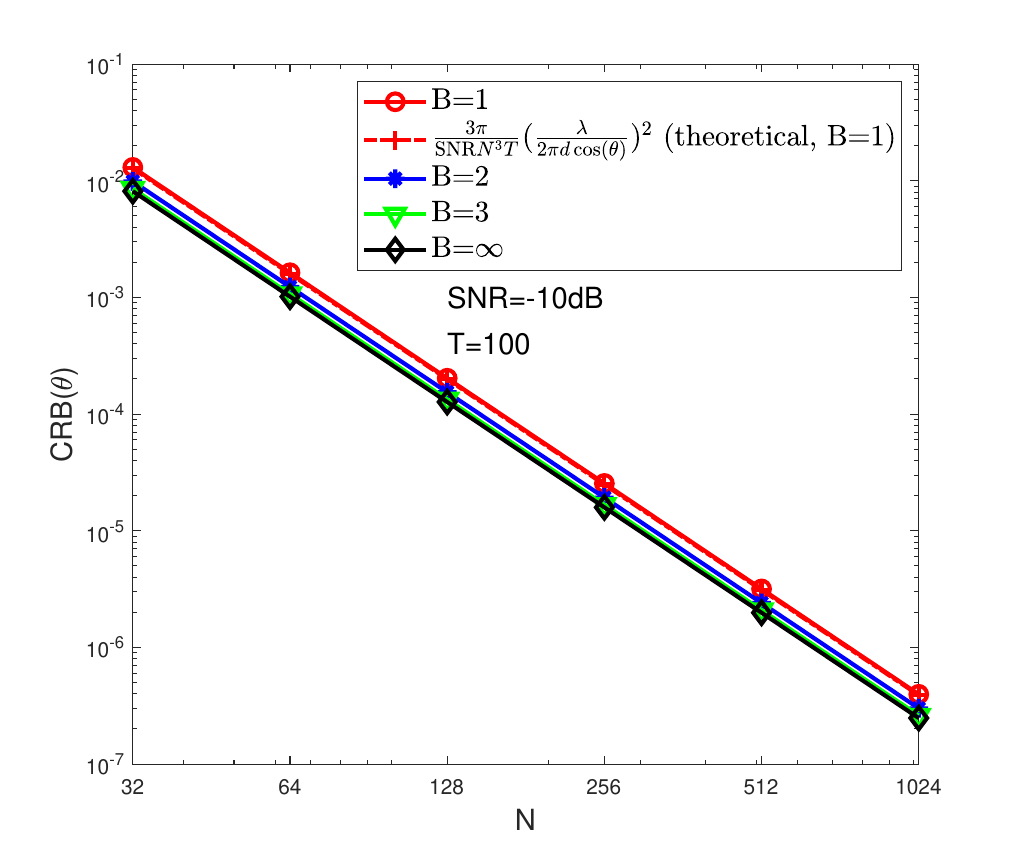}}
    \subfigure[]{
    \label{CRBvsNhighSNR} 
    \includegraphics[width=52mm]{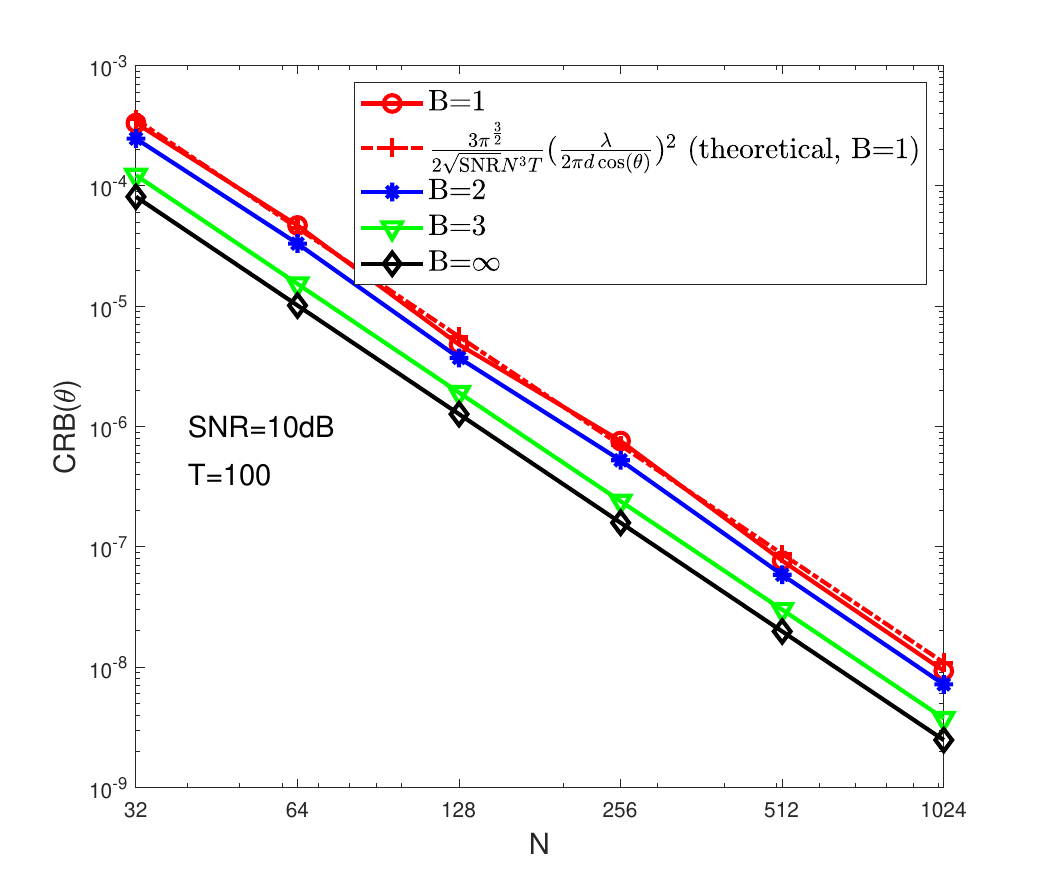}}
  \caption{The CRB of a single source at $30^{\circ}$ and its asymptotic results.}
  \label{CRBasyp} 
\end{figure*}

This experiment is to validate the asymptotic property (\ref{asymELCRB}). Fig. \ref{CRBasyp} shows that the CRB under one bit quantization is inversely proportional to the cubic of the number of measurements $N$, the number of snapshots. For low SNR, the CRB under one bit quantization is inversely proportional to the SNR. While for high SNR, the CRB under one bit quantization is inversely proportional to the square root of the SNR. It can be seen that the asymptotic bound (\ref{asymELCRB}) provides a very good approximation for the true CRB. For the high SNR scenario shown in Fig. \ref{CRBvsThighSNR}, the approximation is not accurate when $T=1$, because our bound is obtained by averaging over the phase of the amplitude of the frequency. As $T$ increases to $10$, (\ref{asymELCRB}) approximates the true CRB well.

\subsection{${{\rm NMSE}}(\hat{\mathbf Z})$ versus the Number of Iteration}
\begin{figure}
  \centering
    \includegraphics[width=2.8in]{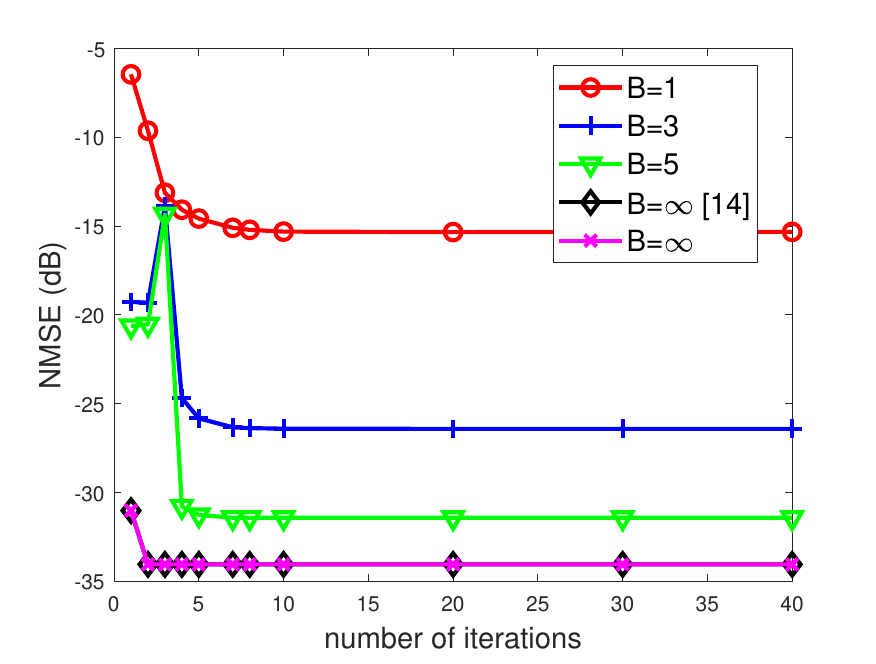}
  \caption{${{\rm NMSE}}(\hat{\mathbf Z})$ versus the number of iteration, here $N=80$, $K=3$, $T=50$.}
  \label{NMSEvsIter} 
\end{figure}

The first experiment evaluates the ${{\rm NMSE}}(\hat{\mathbf Z})$ versus the number of iterations for $B=1,3,5$ and unquantized measurements. The DOAs are fixed as $[-3^{\circ},2^{\circ},75^{\circ}]$, and the SNRs are $[8,16,12]$ dB, respectively. The results are averaged over $50$ MC trials and are presented in Fig. \ref{NMSEvsIter}.  Note that MVALSE-EP converges very quickly. Meanwhile, the NMSE performance of the MVALSE-EP improves as the bit-depth increases.
\subsection{Performance versus SNR}
\begin{figure*}
  \centering
  \subfigure[]{
    \label{NMSEZvsSNRmin} 
    \includegraphics[width=52mm]{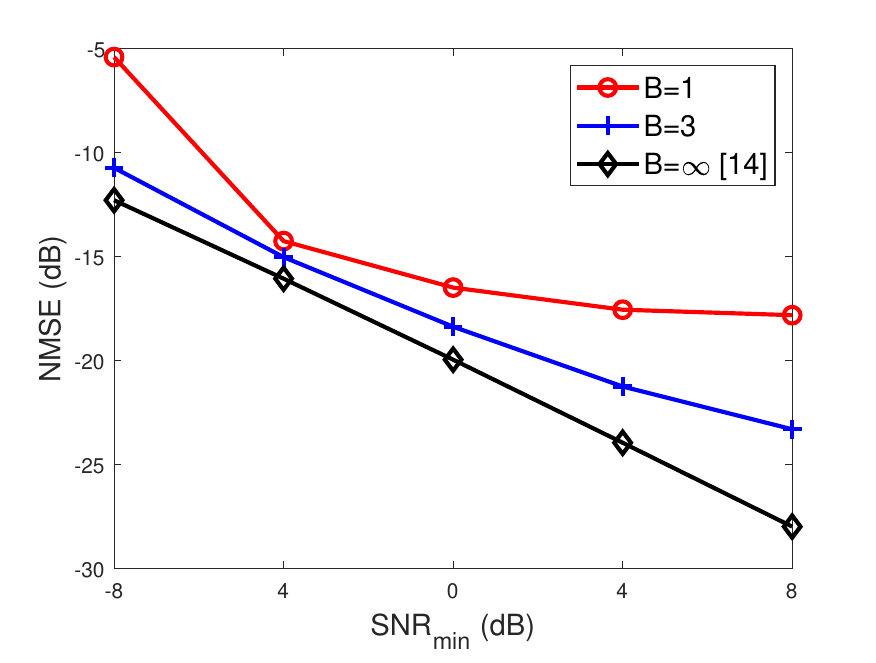}}
  \subfigure[]{
    \label{PrKequalKhatvsSNRmin} 
    \includegraphics[width=52mm]{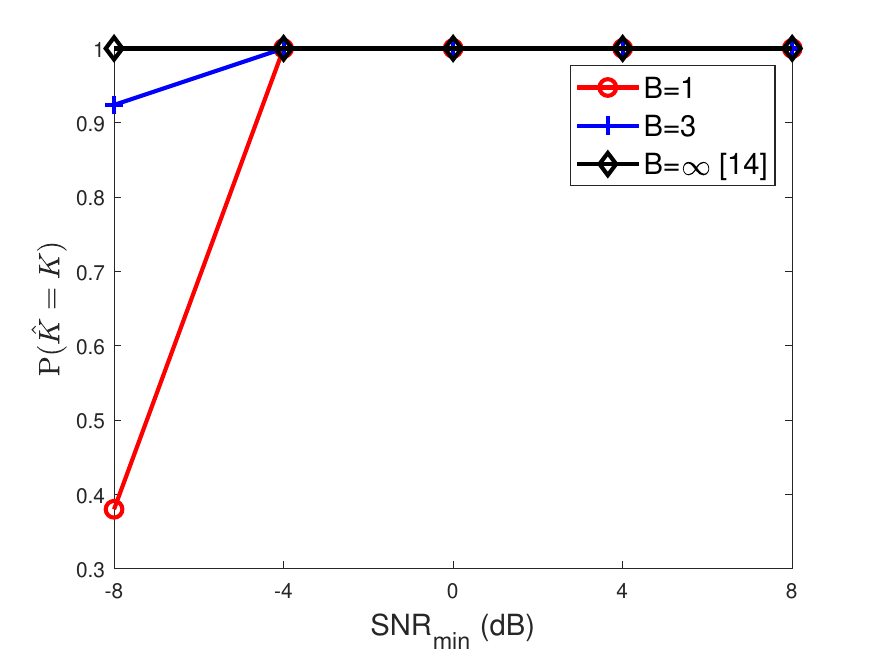}}
  \subfigure[]{
    \label{RMSEvsSNRmin} 
    \includegraphics[width=52mm]{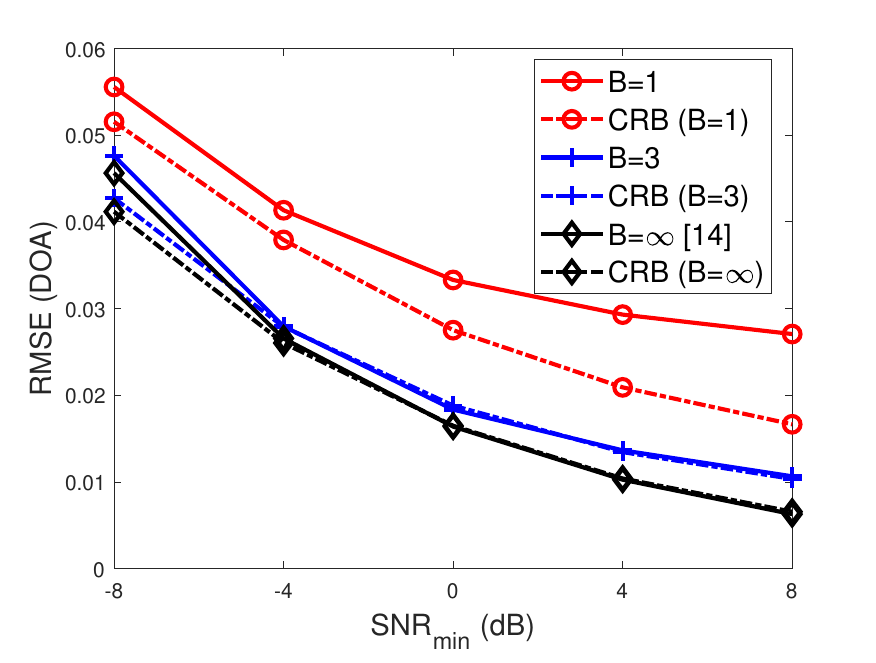}}
  \caption{The performance of the MVALSE-EP versus the minimal SNR ${\rm SNR}_{\rm min}$. Here we set $N=100$, $K=3$, $T=50$ dB, $\Delta=5$ dB, ${\boldsymbol\theta}=[-3^{\circ},2^{\circ},75^{\circ}]$.}
  \label{ResvsSNRmin} 
\end{figure*}

The performance of the MVALSE-EP versus SNR is investigated and the results are plotted in Fig. \ref{ResvsSNRmin}. It can be seen that as ${\rm SNR}_{\rm min}$ increases, the NMSE of the signal decreases and slows down, especially for $1$ bit quantization. The model order estimation probability increases as ${\rm SNR}_{\rm min}$ increases and approaches $1$ for ${\rm SNR}_{\rm min}\geq -4$ dB. For $3$ bit quantization and no quantization, the RMSE of the DOA of the MVALSE-EP approaches to the CRB for ${\rm SNR}_{\rm min}\geq -4$ dB. While for $1$ bit quantization, there always exists a performance gap.
\subsection{Performance versus Number of Snapshots}
\begin{figure*}
  \centering
  \subfigure[]{
    \label{NMSEZvsT} 
    \includegraphics[width=52mm]{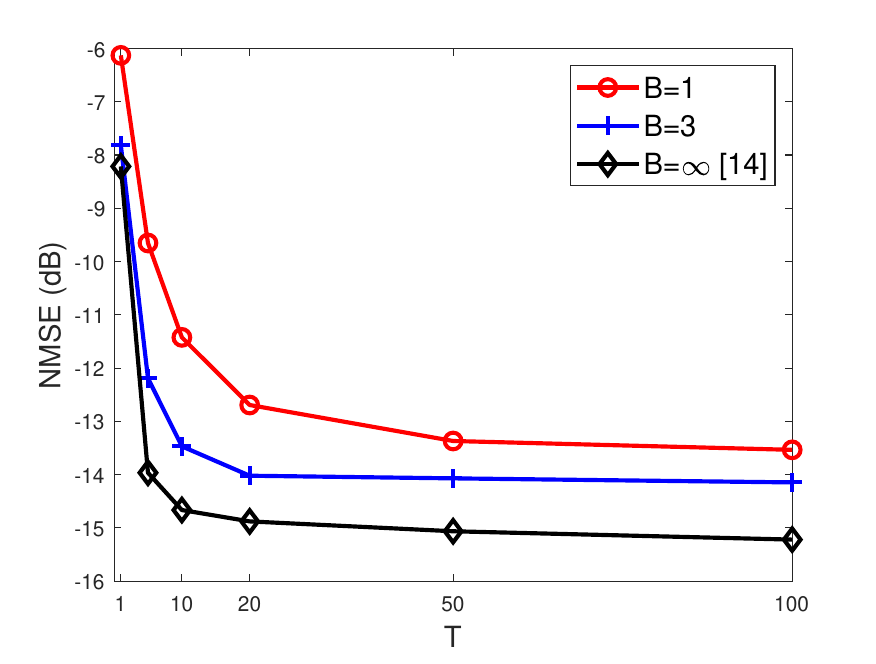}}
  \subfigure[]{
    \label{ProbKhatvsT} 
    \includegraphics[width=52mm]{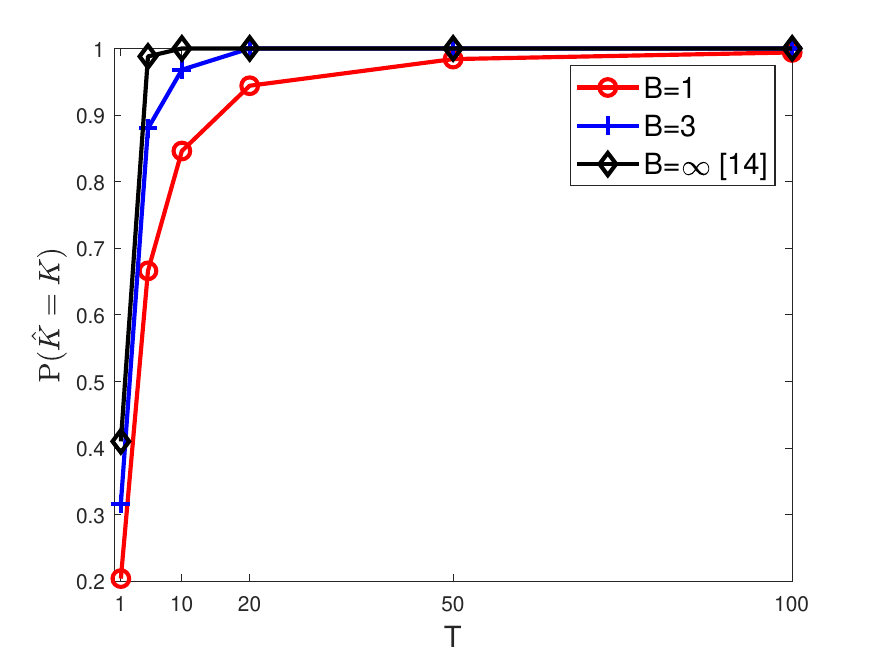}}
  \subfigure[]{
    \label{RMSEvsT} 
    \includegraphics[width=52mm]{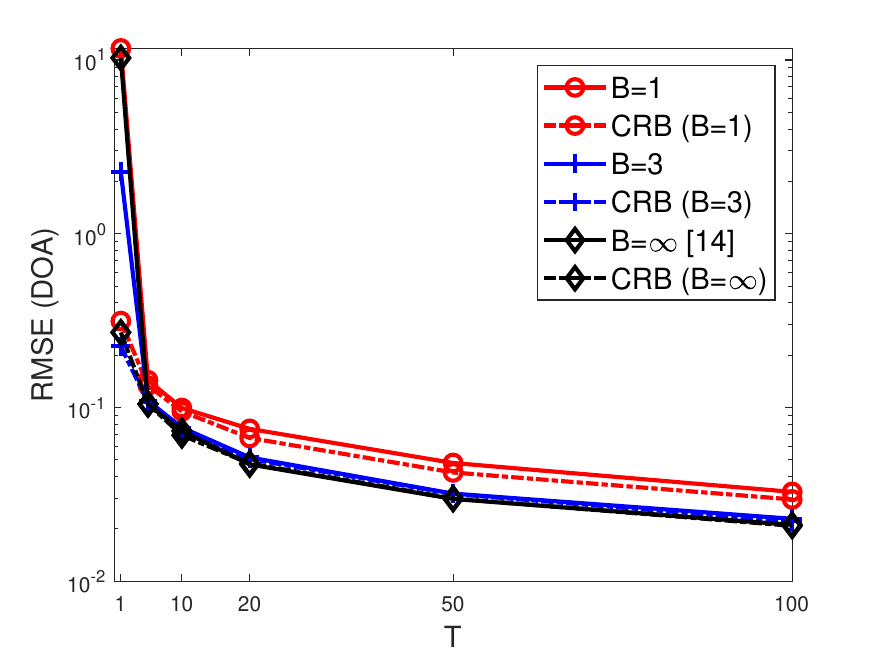}}
  \caption{The performance of the MVALSE-EP versus the number of snapshots $T$. Here we set $N=M=100$, $K=3$, ${\rm SNR}_{\rm min}=-5$ dB, $\Delta=5$ dB, ${\boldsymbol\theta}=[-3^{\circ},2^{\circ},75^{\circ}]$.}
  \label{ResvsT} 
\end{figure*}

The performance of the MVALSE-EP versus the number of snapshots $T$ is investigated and the results are plotted in Fig. \ref{ResvsT}. It can be seen that the NMSE of the signal decreases as the number of snapshots $T$ increases and saturates for $T\geq 50$. The successful model order estimation probability increases as the number of snapshots increases and approaches $1$. As for the RMSE of DOAs, it can be seen that the RMSE is significantly larger than their corresponding CRBs under $1$ bit, $3$ quantization and no quantization for $T\leq 2$. The reason is that when the number of snapshots is small, the number of the estimated DOAs is $3$ while the DOAs are estimated with large error, i.e., some DOAs are not detected while some false alarms are generated. As the number of snapshots increases, the probability of such events decreases and the RMSEs of DOAs approaches to the CRB. This demonstrates that increasing the number of snapshots benefits the DOA estimation.
\subsection{Real Data}
\begin{figure*}
  \centering
  \subfigure[]{
    \label{sce1} 
    \includegraphics[height=35mm]{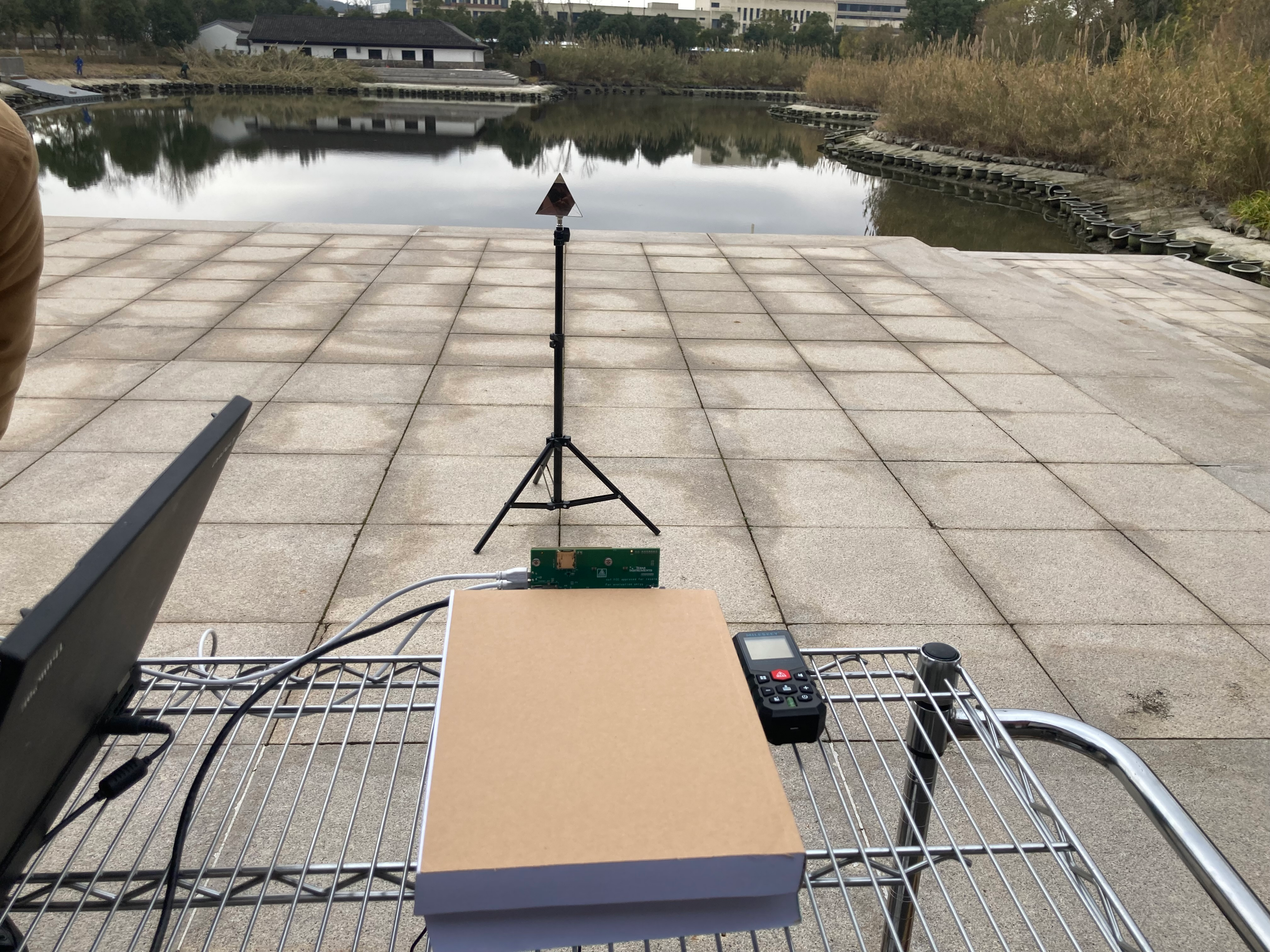}}
  \subfigure[]{
    \label{sce2} 
    \includegraphics[height=35mm]{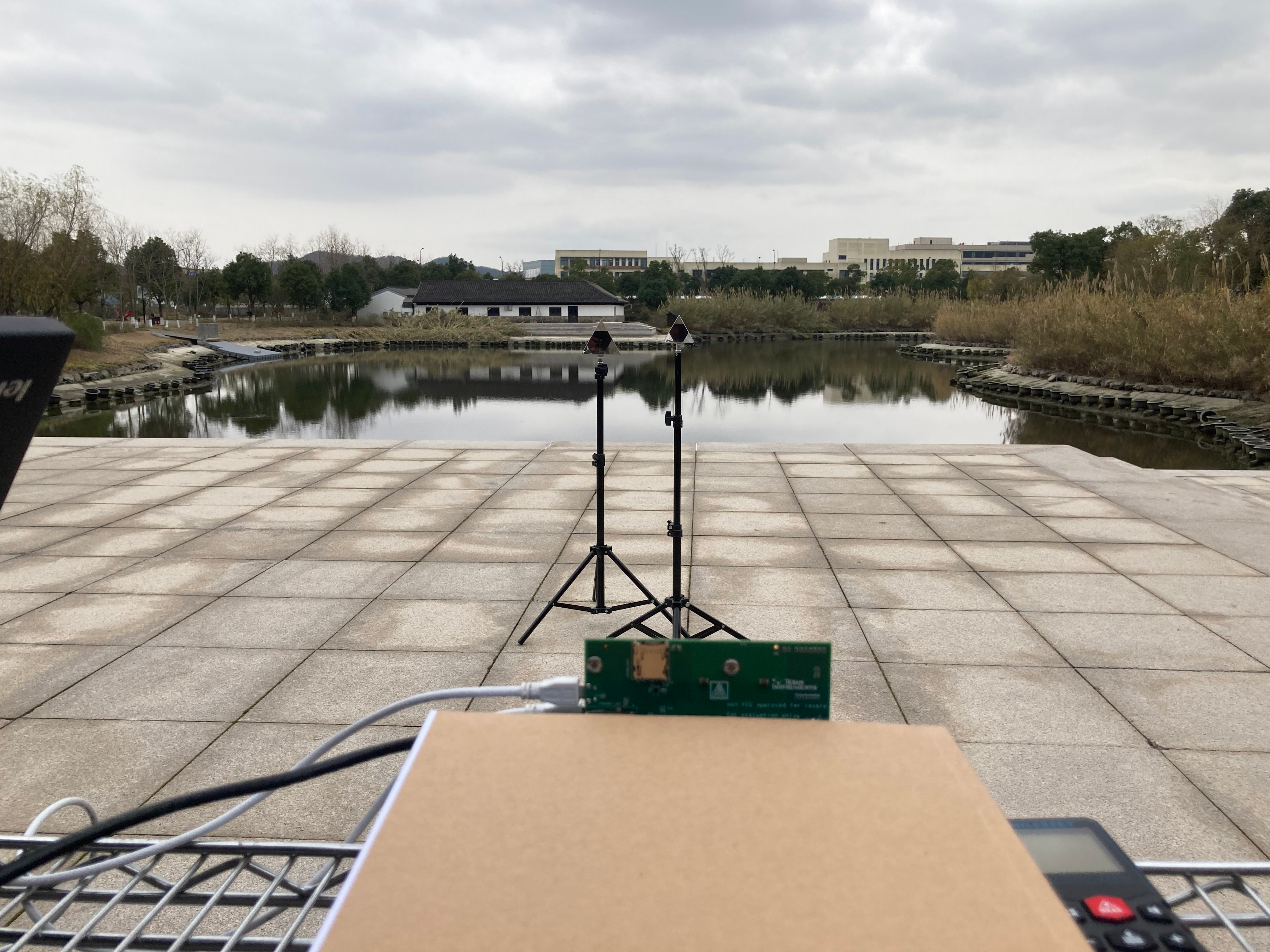}}
    \subfigure[]{
    \label{sce3} 
    \includegraphics[height=35mm]{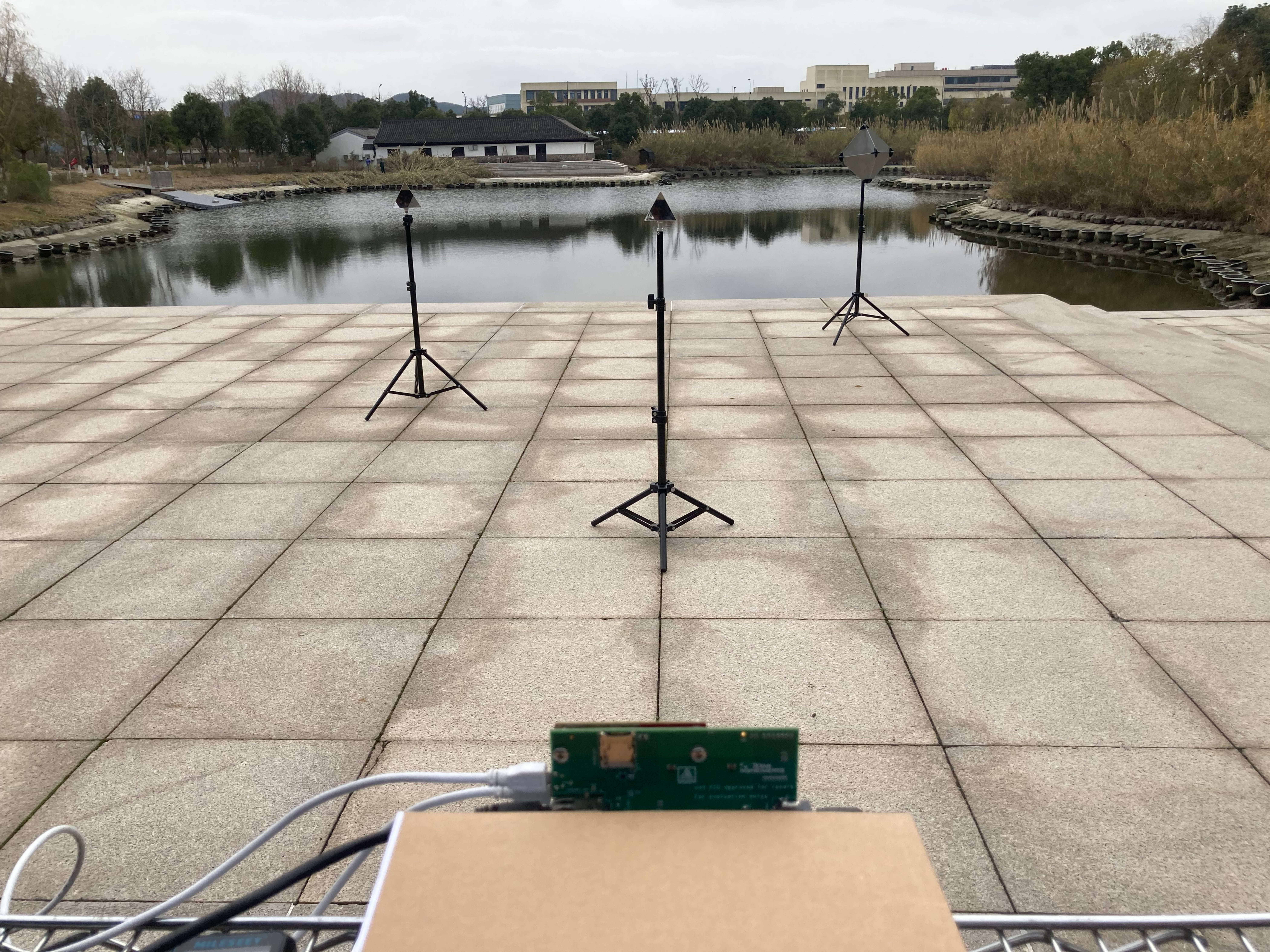}}
  \caption{The three experiments.}
  \label{senarios} 
\end{figure*}

Thus subsection uses the LFMCW AWR1642 Single-Chip to perform the range estimation. The chirp rate is set as $\kappa = 29.982\times 10^{12} {\rm M}/{\rm Hz}^2$, the sampling frequency is $F_s=10$ MHz, and the maximum distance is $r_{\rm max}=\frac{cF_s}{2\kappa}=50$ m where $c=3\times 10^{8}$ denotes the speed of the electromagnetic wave. $\lambda$ (\ref{delta-k-active_tune}) is set as 6 and 15 for 1 bit quantization and 12 bit quantization, respectively. Given that MVALSE-EP outputs the frequencies estimates $\hat{\boldsymbol \omega}$, the range estimates are $\hat{\mathbf r}=\frac{\hat{\boldsymbol \omega}}{2\pi}r_{\rm max}$. The number of fast time samples is $N=128$ and the number of snapshots is $16$. We first put corners as targets and test the algorithm in three experiments, as shown in Fig. \ref{senarios}. For the first experiment in Fig. \ref{sce1}, corner 1 is set just in front of the radar with the measured radial distance being 2.25m. For experiment 2, a little larger corner named corner 2 is added with the measured horizontal and vertical distance being $0.25$m and $2.5$m, and the radial distance can be calculated as $\sqrt{0.25^2+2.5^2}\approx 2.51$m. For experiment 3, the position of corner 1 is kept unchanged, and the measured horizontal and vertical distance of corner 2 is $1$m and $3.75$m, corresponding to the radial distance $3.88$m. A much larger Corner named corner 3 is added with the measured horizontal and vertical distance being $1$m and $5.25$m, and the radial distance can be calculated as $\sqrt{0.25^2+2.5^2}\approx 5.34$m.

\begin{figure}[h!t]
\centering
\includegraphics[width=3.8in]{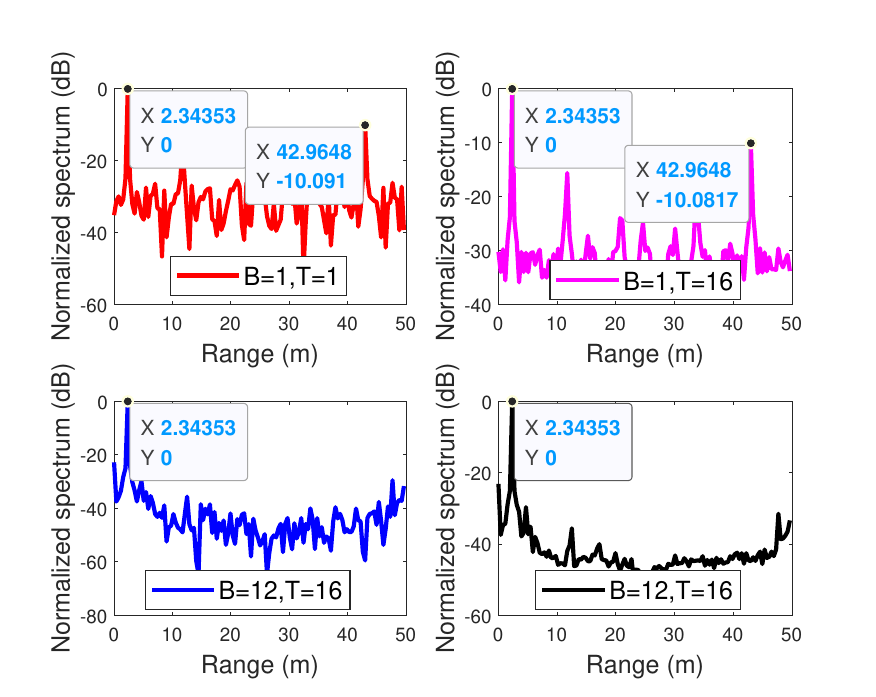}
\caption{The normalized spectrum of the 1 bit and 12 bit signal for experiment 1.}
\label{sce1tarFFT}                                                                                                                                                                                                                                                       \end{figure}

For experiment 1, the normalized spectrum and the results are shown in Fig. \ref{sce1tarFFT} and Table \ref{rangestsce1}, respectively. It can be seen that for both 1 bit and 12 bit quantization, the highest peak occurs at $2.34$m. For 1 bit quantization, the second peak occurs at $42.96\approx 50-2.34\times 3=42.98$m, corresponding to the third harmonic and is $10$ dB lower than the highest peak. For 1 bit quantization, MVALSE-EP estimates the range of corner 1 as $2.32$ m for the single snapshot and multiple snapshots. While for 12 bit quantization, the number of detected targets is $3$, and MVALSE-EP still estimates the range of corner 1 as $2.32$ m for the single snapshot and multiple snapshots.

\begin{table*}[h!t]
\centering
\caption{The range estimation performance of the MVALSE-EP for experiment 1.}
\label{rangestsce1}
\begin{tabular}{|c|c|c|}
  \hline
  B \textbackslash T& T=1 (range (m), Amp. (dB))&T=16 (range (m), Amp. (dB))\\\hline
  1 bit & (2.32,)  & (2.32,) \\\hline
  12 bit & (2.32,46.9),(0.08,23.6), (2.46, 26.4) & (2.32, 46.6), (0.09, 23.3),(2.44, 27.3) \\
  \hline
\end{tabular}
\end{table*}

\begin{figure}[h!t]
\centering
\includegraphics[width=3.8in]{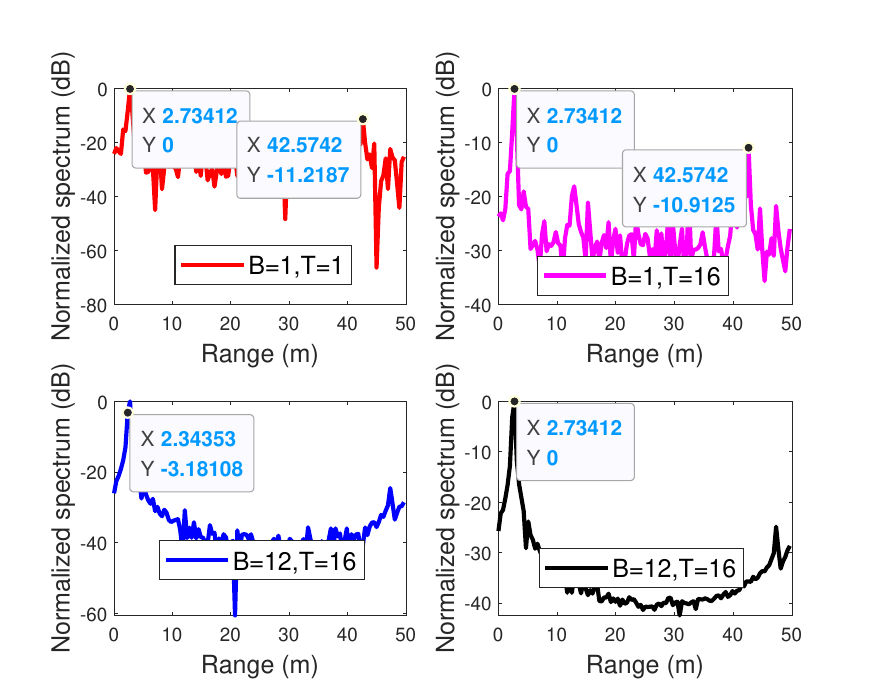}
\caption{The normalized spectrum of the 1 bit and 12 bit signal for experiment 2.}
\label{sce2tarFFT}                                                                                                                                                                                                                                                       \end{figure}

\begin{table*}[h!t]
\centering
\caption{The range estimation performance of the MVALSE-EP for experiment 2.}
\label{rangestsce2}
\begin{tabular}{|c|c|c|}
  \hline
  B \textbackslash T& T=1 (range (m), Amp. (dB))&T=16 (range (m), Amp. (dB))\\\hline
  1 bit & (2.65,0),(2.21,-6.21)  & (2.66,0),(2.21,-5.99) \\\hline
  12 bit & (2.61,50.0),(2.31,46.4), (0.08, 23.7),(1.31, 22.6) & (2.61, 50.0), (2.31, 46.3),(0.11, 24.0) \\
  \hline
\end{tabular}
\end{table*}

For experiment 2, Fig. \ref{sce2tarFFT} shows that FFT approach can not resolve the two targets. Table \ref{rangestsce2} show that MVALSE-EP detects the two corners and estimates the ranges. Besides, the amplitude of the first corner is $46$ dB, consistent with the results obtained in scenario 1. Also, the amplitude of corner 1 is about 4 dB lower than the second corner.

\begin{figure}[h!t]
\centering
\includegraphics[width=3.8in]{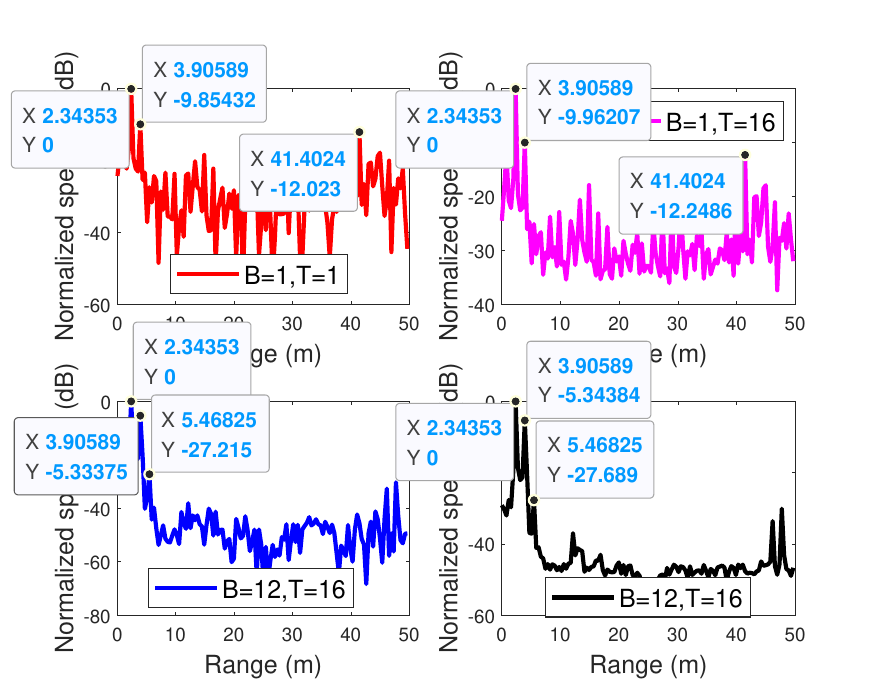}
\caption{The normalized spectrum of the 1 bit and 12 bit signal for experiment 3.}
\label{sce3tarFFT}                                                                                                                                                                                                                                                       \end{figure}

For experiment 3, the spectrum in Fig. \ref{sce3tarFFT} shows that after 1 bit quantization, the peak corresponding to corner 3 disappears. For the reconstruction results shown in Table \ref{rangestsce3}, MVALSE-EP only detects the strongest target corresponding to corner 1 with a single snapshot. For multiple snapshots, corner 2 is also detected and its amplitude is $6.68$ dB lower than corner 1. For high resolution data, MVALSE detects the three corners, and also the leakage component whose amplitude is comparable to corner $3$. This demonstrates that under 1 bit quantization, algorithms can not detect the weak target in the presence of the strong target, where the weak target (about $23$ dB) is $25$ dB lower than the strongest target (about $48.6$ dB).

\begin{table*}[h!t]
\centering
\caption{The range estimation performance of the MVALSE-EP for experiment 3.}
\label{rangestsce3}
\begin{tabular}{|c|c|c|}
  \hline
  B \textbackslash T& T=1 (range (m), Amp. (dB))&T=16 (range (m), Amp. (dB))\\\hline
  1 bit & (2.29,0)  & (2.29,0),(3.97,-6.68) \\\hline
  12 bit & (2.30,48.6),(3.97,43.3), (0.12, 22.9),(5.44, 23.0) & (2.30, 48.6), (3.97, 43.2),(0.12, 23.0),(5.41, 22.7) \\
  \hline
\end{tabular}
\end{table*}
\begin{figure*}
  \centering
  \subfigure[]{
    \label{sce4} 
    \includegraphics[height=53mm]{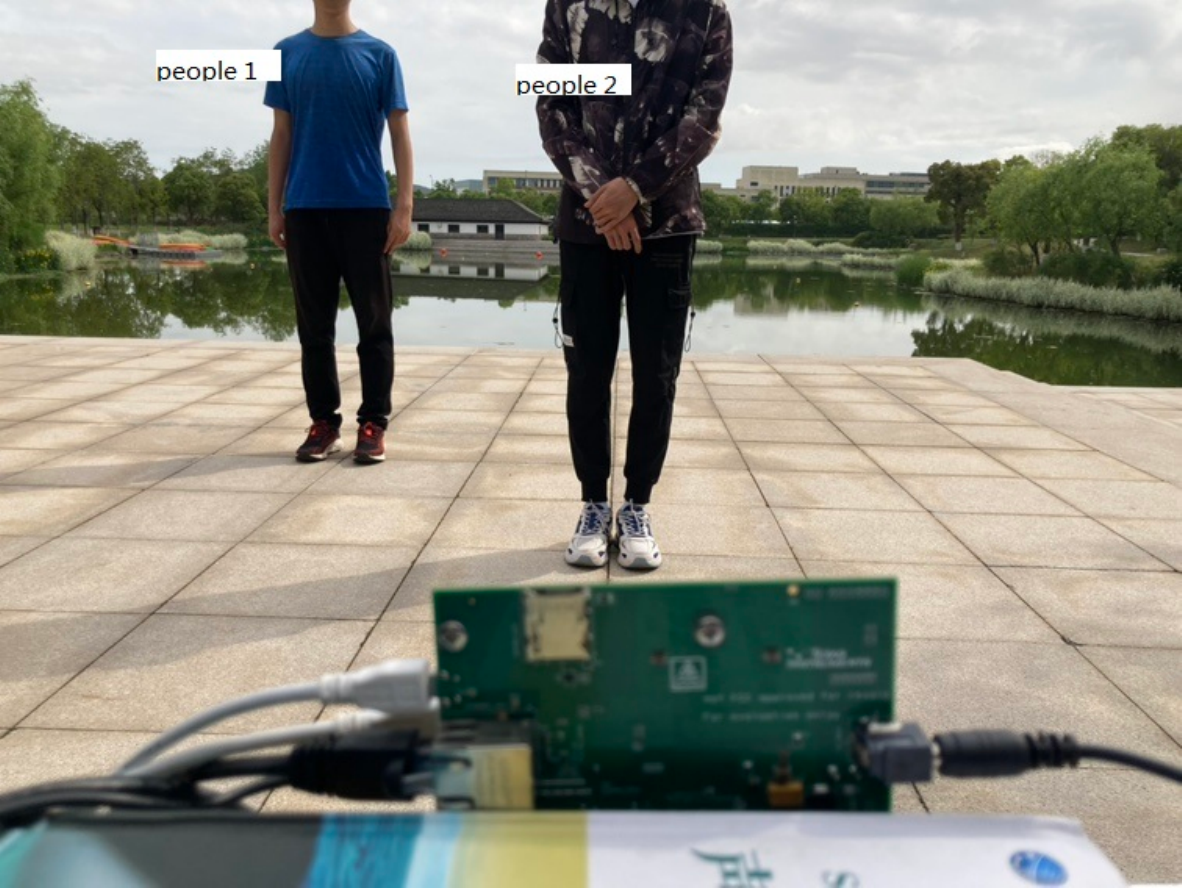}}
  \subfigure[]{
    \label{sce5} 
    \includegraphics[height=53mm]{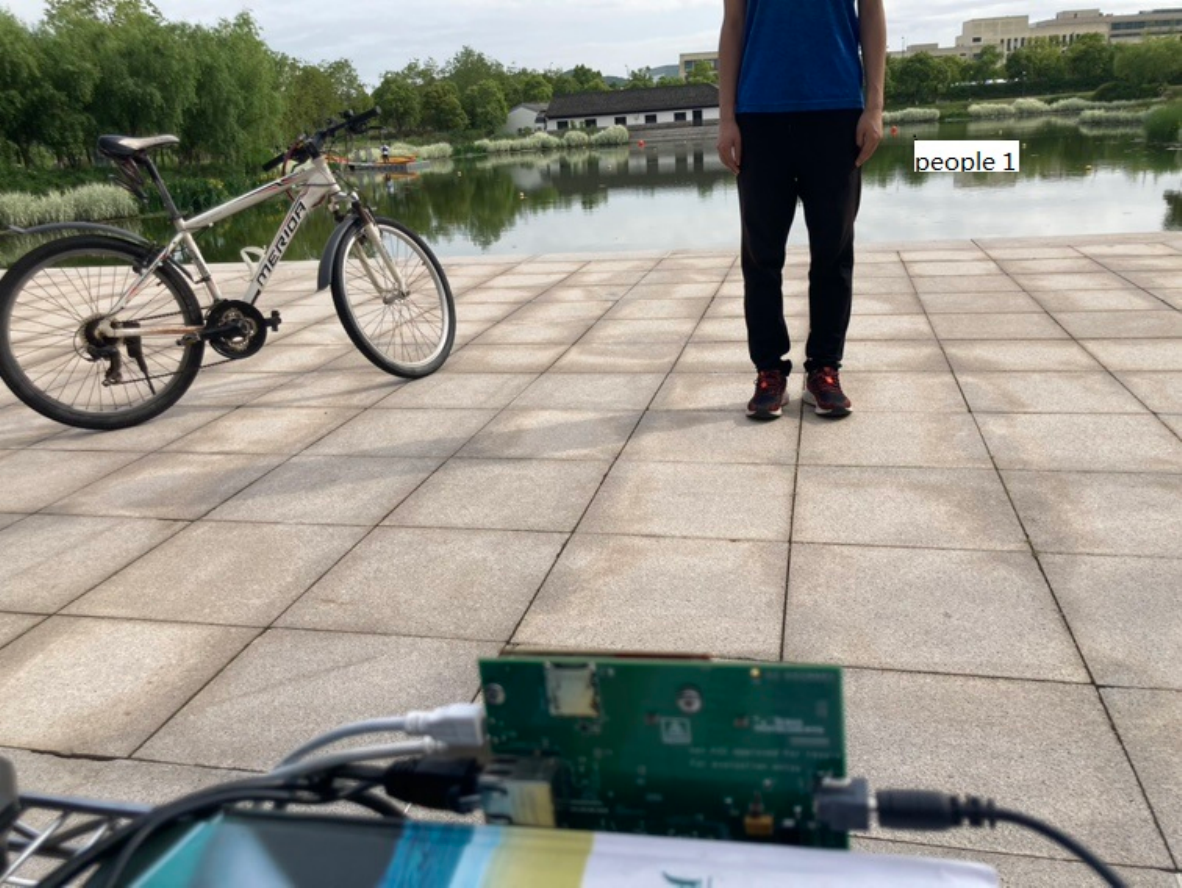}}
  \caption{Two people experiment and one people, one bicycle experiment.}
  \label{senariospeople} 
\end{figure*}
\begin{table*}[h!t]
\centering
\caption{The range estimation performance of the MVALSE-EP for experiment 4.}
\label{rangestsce4}
\begin{tabular}{|c|c|c|}
  \hline
  B \textbackslash T& T=1 (range (m), Amp. (dB))&T=16 (range (m), Amp. (dB))\\\hline
  1 bit & (3.42,0),(2.35,-4.9),(0.13,-5.7)  & (3.42,0),(2.40,-6.58) \\\hline
  12 bit & (3.46,27.4),(2.32,21.2), (3.77, 21.7),(0.12, 17.8) & (3.46, 27.2), (3.78, 21.7),(2.36, 20.4),(0.12, 18.4) \\
  \hline
\end{tabular}
\end{table*}

\begin{table*}[h!t]
\centering
\caption{The range estimation performance of the MVALSE-EP for experiment 5.}
\label{rangestsce5}
\begin{tabular}{|c|c|c|}
  \hline
  B \textbackslash T& T=1 (range (m), Amp. (dB))&T=16 (range (m), Amp. (dB))\\\hline
  1 bit & (3.55,0),(2.75,-1.7)  & (3.55,0),(2.74,-1.7) \\\hline
  12 bit & (3.56,30.7),(2.77,28.1), (0.08, 21.5),(4.02, 18.8) & (3.56, 30.5), (2.76, 28.8),(0.08, 22.0),(4.00, 16.9) \\
  \hline
\end{tabular}
\end{table*}
For the fourth experiment shown in Fig. \ref{sce4}, the ranges of people 1 and people 2 are 3.42 m and 2.27 m, respectively. The reconstruction results are shown in Table \ref{rangestsce4}. It can be seen that even under 1 bit quantization, MVALSE-EP detects two people and outputs the range estimates as $3.42$m and $2.35$m for single snapshot case, $3.42$m and $2.40$m for multiple snapshot case. Besides, the amplitude of people 1 and people 2 are about 27 dB and 21 dB, respectively.

For the last experiment, the ranges of the bicycle and people 1 are 3.41m and 2.77m, respectively. Results are shown in Table \ref{rangestsce5}. It can be seen that  the bicycle and people 1 are detected under 1 bit and 12 bit quantization. Besides, the amplitudes of the bicycle and people 1 are about 31 dB and 28 dB, respectively. Note that  the amplitudes of people 1 is consistent with the results in experiment 4.
\section{Conclusion}\label{con}
In this paper, the multisnapshot line spectral estimation problem from coarsely quantized measurements is studied. The CRB is derived and the effects of the system parameters such as SNR, number of measurements and snapshots on the estimation accuracy are revealed in a single frequency scenario. Then, an MVALSE-EP algorithm which automatically estimates the frequencies, the model order and noise variance is proposed. Substantial numerical experiments including real data set are conducted to show the effectiveness of the MVALSE-EP.
\section{Appendix}
\subsection{CRB for the general case}\label{generalFIM}
Note that
\begin{align}\label{derpartres}
\frac{\partial {\boldsymbol \mu}({\boldsymbol \kappa})}{\partial {\boldsymbol \kappa}^{\rm T}}=
\begin{pmatrix}
  \bar{\mathbf A} &  & &  &  \bar{\mathbf D}(1) \\
   & \bar{\mathbf A} & &   & \bar{\mathbf D}(2) \\
   &  & \ddots &   & \vdots  \\
   & &  & \bar{\mathbf A}& \bar{\mathbf D}(T)  \end{pmatrix},
\end{align}
where $\bar{\mathbf A}$ and $\bar{\mathbf D}(t)$ are defined in (\ref{Abar}) and (\ref{Dbart}), respectively.

Substituting (\ref{derpartres}) in (\ref{FIMcal}) yields
\begin{align}
{\mathbf I}({\boldsymbol \kappa})=\frac{2}{\sigma^2}
\begin{pmatrix}
  \bar{\mathbf H}(1) &  & &  & \bar{\boldsymbol \Delta}(1) \\
   & \bar{\mathbf H}(2) & &   & \bar{\boldsymbol \Delta}(2) \\
   &  & \ddots &   & \vdots  \\
   & &  & \bar{\mathbf H}(T)& \bar{\boldsymbol \Delta}(T)  \\
  \bar{\boldsymbol \Delta}^{\rm T}(1) & \bar{\boldsymbol \Delta}^{\rm T}(2) &\cdots & \bar{\boldsymbol \Delta}^{\rm T}(T) & \bar{\boldsymbol \Gamma} \\
\end{pmatrix},
\end{align}
where $\bar{\mathbf H}(t)$ and $\bar{\boldsymbol \Delta}(t)$ are defined in (\ref{Hbart}) and (\ref{Deltabart}), respectively, $\bar{\boldsymbol \Gamma}$ is
\begin{align}
\bar{\boldsymbol \Gamma}= \sum\limits_{t=1}^T\bar{\mathbf D}^{\rm T}(t){\boldsymbol\Lambda}(t)\bar{\mathbf D}(t).
\end{align}
Note that for the inverse of a block matrix, one has
\begin{align}
\left[
  \begin{array}{cc}
    {\mathbf A} & {\mathbf B} \\
    {\mathbf B}^{\rm T} & {\mathbf C} \\
  \end{array}
\right]^{-1}=
\left[
  \begin{array}{cc}
    {\mathbf A}^{-1}+{\mathbf A}^{-1}{\mathbf B}{\mathbf S}^{-1}{\mathbf B}^{\rm T}{\mathbf A}^{-1} & -{\mathbf A}^{-1}{\mathbf B}{\mathbf S}^{-1} \\
    -{\mathbf S}^{-1}{\mathbf B}^{\rm T}{\mathbf A}^{-1} & {\mathbf S}^{-1} \\
  \end{array}
\right],
\end{align}
where ${\mathbf S}={\mathbf C}-{\mathbf B}^{\rm T}{\mathbf A}^{-1}{\mathbf B}$. Therefore, ${\rm CRB}({\boldsymbol \omega})$ is
\begin{align}
{\rm CRB}({\boldsymbol \omega})=\frac{\sigma^2}{2}\left(\bar{\boldsymbol \Gamma}-[\bar{\boldsymbol \Delta}^{\rm T}(1),\bar{\boldsymbol \Delta}^{\rm T}(2),\cdots,\bar{\boldsymbol \Delta}^{\rm T}(T)]\left[\begin{matrix}
  \bar{\mathbf H}(1) &  & &   \\
   & \bar{\mathbf H}(2) & &    \\
   &  & \ddots &   &   \\
   & &  & \bar{\mathbf H}(T)   \\
\end{matrix}\right]^{-1}\left[\begin{array}{c}
               \bar{\boldsymbol \Delta}({1}) \\
               \bar{\boldsymbol \Delta}({2}) \\
               \vdots \\
               \bar{\boldsymbol \Delta}({T})
             \end{array}\right]
\right)^{-1},\notag
\end{align}
and performing further simplification yields (\ref{CRBinv}).
\bibliographystyle{IEEEbib}
\bibliography{strings,refs}

\end{document}